\documentclass[twocolumn]{aastex631}

\newcommand\aastex{AAS\TeX}

\usepackage{comment}
\usepackage{mhchem}
\usepackage[flushleft]{threeparttablex}
\usepackage{tablefootnote}

\newcommand \prodimo{\texttt{ProDiMo}}
\newcommand \Go {\chi}
\newcommand \Mg {M_\mathrm{g}}
\newcommand \Md {M_\mathrm{d}}
\newcommand \Rtap {R_\mathrm{tap}}
\newcommand \nspectral {n_\mathrm{13-25}}
\newcommand \Mearth {M_\oplus}
\newcommand \duck {\texttt{DuCK}}
\newcommand \pyroxene {\mathrm{Mg}_{0.7}\mathrm{Fe}_{0.3}\mathrm{SiO}_{3}}
\newcommand \olivine {\mathrm{Mg}\mathrm{Fe}\mathrm{SiO}_{4}}
\newcommand \forsterite {\mathrm{Mg}_{2}\mathrm{SiO}_{4}}
\newcommand \enstatite {\mathrm{Mg}_{0.96}\mathrm{Fe}_{0.04}\mathrm{SiO}_{3}}
\newcommand \fayalite {\mathrm{Fe}_{2}\mathrm{SiO}_{4}}
\newcommand \silica {\mathrm{SiO}_{2}}
\newcommand \Tave {\langle T_\mathrm{gas} \rangle}
\newcommand \Nave {\langle N \rangle}
\newcommand \Rave {\langle R_\mathrm{eff} \rangle}
\newcommand \torus {\texttt{TORUS-3DPDR}}
\newcommand{\rt}[1]{{#1}}

\accepted{for publication in ApJ, March 28, 2025}

\begin{document}

\title{XUE. Thermochemical Modeling Suggests a Compact and Gas-Depleted Structure for a Distant, Irradiated Protoplanetary Disk}

\correspondingauthor{Bayron Portilla-Revelo}
\email{bmp5924@psu.edu}

\author[0000-0002-6278-9006]{Bayron Portilla-Revelo}
\affiliation{Department of Astronomy and Astrophysics, The Pennsylvania State University, 525 Davey Laboratory, University Park, PA 16802, USA}
\affiliation{Center for Exoplanets and Habitable Worlds, Penn State University, 525 Davey Laboratory, 251 Pollock Road, University Park, PA, 16802, USA.}

\author[0000-0002-6137-8280]{Konstantin V. Getman}
\affiliation{Department of Astronomy and Astrophysics, The Pennsylvania State University, 525 Davey Laboratory, University Park, PA 16802, USA}

\author[0000-0001-9698-4080]{María Claudia Ramírez-Tannus}
\affiliation{Max-Planck Institut für Astronomie (MPIA), Königstuhl 17, D-69117 Heidelberg, Germany}

\author[0000-0002-9593-7618]{Thomas J. Haworth}
\affiliation{Astronomy Unit, School of Physics and Astronomy, Queen Mary University of London, London E1 4NS, UK}

\author[0000-0002-5462-9387]{Rens Waters}
\affiliation{Department of Astrophysics/IMAPP, Radboud University, PO Box 9010, 6500 GL Nijmegen, The Netherlands}
\affiliation{SRON Netherlands Institute for Space Research, Niels Bohrweg 4, NL-2333 CA Leiden, the Netherlands}

\author[0000-0001-8068-0891]{Arjan Bik}
\affiliation{Department of Astronomy, Stockholm University, AlbaNova University Center, SE-10691 Stockholm, Sweden}

\author[0000-0002-5077-6734]{Eric D. Feigelson}
\affiliation{Department of Astronomy and Astrophysics, The Pennsylvania State University, 525 Davey Laboratory, University Park, PA 16802, USA}
\affiliation{Center for Astrostatistics, Pennsylvania State University, PA, USA}

\author[0000-0001-7455-5349]{Inga Kamp}
\affiliation{Kapteyn Astronomical Institute, University of Groningen, PO BOX 800, 9700 AV Groningen, The Netherlands}

\author[0000-0002-1284-5831]{Sierk E. van Terwisga}
\affiliation{Space Research Institute, Austrian Academy of Sciences, Schmiedlstr. 6, 8042, Graz, Austria}

\author[0009-0003-7663-5280]{Jenny Frediani}
\affiliation{Department of Astronomy, Stockholm University, AlbaNova University Center, SE-10691 Stockholm, Sweden}

\author[0000-0002-1493-300X]{Thomas Henning}
\affiliation{Max-Planck Institut für Astronomie (MPIA), Königstuhl 17, D-69117 Heidelberg, Germany}

\author[0000-0002-7501-9801]{Andrew J. Winter}
\affiliation{Université Côte d’Azur, Observatoire de la Côte d’Azur, CNRS, Laboratoire Lagrange, 06300 Nice, France}
\affiliation{Max-Planck Institut für Astronomie (MPIA), Königstuhl 17, D-69117 Heidelberg, Germany}

\author[0000-0002-4650-594X]{Veronica Roccatagliata}
\affiliation{Dipartimento di Fisica e Astronomia, Alma Mater Studiorum, Universitàdi Bologna,  Via Gobetti 93/2, I-40129 Bologna,  Italy}
\affiliation{INAF-Osservatorio Astrofisico di Arcetri,  Largo E. Fermi 5, I-50125 Firenze,  Italy}

\author[0000-0003-3130-7796]{Thomas Preibisch}
\affiliation{Universit{\"a}ts-Sternwarte M{\"u}nchen, Ludwig-Maximilians-Universit{\"a}t, Scheinerstr.~1, 81679  M{\"u}nchen, Germany}

\author[0000-0003-2954-7643]{E. Sabbi}
\affiliation{Gemini Observatory/NSFs NOIRLab, 950 N. Cherry Ave., Tucson, AZ 85719, USA}

\author[0000-0002-6091-7924]{Peter Zeidler}
\affiliation{AURA for the European Space Agency (ESA), ESA Office, Space Telescope Science Institute, 3700 San Martin Drive, Baltimore, MD 21218, USA}

\author[0000-0002-0631-7514]{Michael A. Kuhn}
\affiliation{Centre for Astrophysics Research, Department of Physics, Astronomy and Mathematics, University of Hertfordshire, Hatfield, AL10 9AB, UK}

\begin{abstract}
Unveiling the physical structure of protoplanetary disk is crucial for interpreting the diversity of the exoplanet population. Until recently, the census of the physical properties of protoplanetary disks probed by mid-infrared observations was limited to the solar neighborhood ($d \lesssim 250$~pc); however, nearby star-forming regions (SFRs) such as Taurus---where no O-type stars reside---are not representative of the environments where the majority of the planet formation occurs in the Galaxy. The James Webb Space Telescope (JWST) now enables observations of disks in distant high-mass SFRs, where strong external Far-Ultraviolet (FUV) radiation is expected to impact those disks.  Nevertheless, a detailed characterization of externally irradiated disks is still lacking. We use the thermochemical code \prodimo\ to model JWST/MIRI spectroscopy and archival visual/near-infrared photometry aiming to constrain the physical structure of the irradiated disk around the solar-mass star XUE~1 in NGC~6357 ($d \approx 1690$~pc). Our findings are: (1) Mid-infrared dust emission features are explained by amorphous and crystalline silicates with compositions similar to nearby disks. (2) The molecular features detected with MIRI originate within the first $\sim 1$~au, consistent with slab models' results. (3) Our model favors a disk truncated at $10$~au with a gas-to-dust ratio of unity in the outskirts. (4) Comparing models of the same disk structure under different irradiation levels, we find that strong external irradiation raises gas temperature tenfold and boosts water abundance beyond 10~au by a factor of 100. Our findings suggest the inner disk resists external irradiation, retaining the elements necessary for planet formation.

\end{abstract}



\section{Introduction} 
\label{sec:intro}

Protoplanetary disks are a by-product of the stellar formation process and are commonly observed during the first several million years of stellar evolution \citep{Richert2018,Manara2023}. The most fundamental properties of a disk---its size, the amount of dust and gas, as well as its chemical composition---will determine the conditions for planet formation within the disk \citep{Mordasini2012}. Such properties can be altered by external environmental conditions. Three major effects capable of shaping disk evolution that are set by the environment are: star-disk gravitational interactions (e.g., \citealt{Vincke2018}), late stage infall of material from the parent cloud (e.g., \citealt{Padoan2005,Kuffmeier2023,Winter2024}), and external photoevaporation driven by Far-Ultraviolet (FUV) photons (e.g., \citealt{Johnstone1998}). This work focuses on the latter effect.  

Just as OB stars produce ionized H II regions in their natal molecular clouds, they will ionize and photoevaporate the protoplanetary disks of nearby low-mass stars. The external FUV flux measured at a disk's surface scales with the number density of OB stars in the field, which positively correlates with the total number of members in the star-forming region (SFR) \citep{Winter2022}. This implies that disks formed in high-mass SFRs overall experience a substantially higher FUV flux compared to those in low-mass SFRs. Given that at least 50\% of stars and planetary systems form in such massive regions (e.g., \citealt{Krumholz2019}), it follows that planet formation predominantly occurs in environments with strong external FUV fields.

The effect of a strong FUV field on a disk is twofold: it affects the disk's surface density (e.g., \citealt{Armitage2020}), and it alters the disk's chemistry (see e.g., \citealt{Adams2010} for a discussion on the chemical signatures of strong radiation fields in forming planetary systems, and \citealt{Desch2024} on the implications for meteoritics). External photoevaporative winds enhance the rate of gas mass loss leading to an accelerated shrinkage of the disk radius \citep{Haworth2018,Winter2022, coleman2022}. External photoevaporation can also influence the distribution of solids, particularly that of sub-micron-sized grains---which are well coupled to the gas---which will be carried away by the wind \citep{Facchini2016, Sellek2020, Winter2022}. Overall, an accelerated gas mass loss rate will reduce the mass budget and the time available for planet formation and migration (\citealt{Winter2022,Qiao2023,Huang2024}). On the other hand, external FUV radiation is expected to substantially increase the temperature at intermediate and surface layers \citep{Ercolano2022}. A high flux of FUV photons is also expected to cause a different chemical stratification in the disk, leading to significant differences in both the abundance of certain molecular species and their observational signatures, compared to those in non-irradiated disks \citep{Walsh2013,Antonellini2015, Keyte2025}.    

The Orion Molecular Cloud Complex has been one suitable testbed for these theoretical predictions. Since the first observation of irradiated disks \citep{O'dell1993}, multiple works have targeted different sub-regions in the Orion Complex. These studies have allowed to: constrain the time span that some disks have been exposed to UV radiation \citep{Bally1998,Clarke2007}; establish that inner disks around YSO's can survive even in the presence of neighboring ionizing stars \citep{Richert2015}; find evidence for disk mass gradients relative to the distance to the ionizing sources \citep{vanTerwisga2019,vanTerwisga2023}; and compare the cold gas-phase chemistry between irradiated and non-irradiated disks \citep{Diaz-Berrios2024}. 

\rt{More recently, the James Webb Space Telescope (JWST) has observed disks in the Orion Nebula region. \cite{Berne2023} reported the first detection of \ce{CH3+} --- a key molecule in gas-phase organic chemistry --- in the outskirts of the irradiated protoplanetary disk d203-506 (see also \citealt{Zannese2025}). The physical structure of this disk was later characterized by \cite{Berne2024} through modeling of observed near-IR \ce{H2} emission lines. \cite{Zannese2024} conducted a combined near- and mid-infrared spectroscopic study of \ce{OH} in the same disk, revealing a ``warm water cycle'' in which \ce{H2O} is efficiently destroyed and reformed under FUV radiation. Finally, \cite{Goicoechea2024} reported near-IR C I fluorescent emission, triggered by FUV pumping, tracing both the disk's upper layers and the photoevaporative wind.} 


\rt{These findings underscore the importance of studying disks in irradiated environments}. However, one distinctive feature of the Orion Nebular Cluster is that a single, centrally-localized star---$\theta^1$ Ori C---is already responsible for $\sim 80\%$ of the H-ionizing photons. Another region of interest is the OB association IC 1795. Synergistic X-ray and infrared observations allowed to determine both the spatial distribution of disks and the disk fraction dependence on the stellar mass \citep{Roccatagliata2011}. IC 1795 hosts only two O type stars once more.    

The advent of the James Webb Space Telescope (JWST) has opened a new window in the study of irradiated disks. Its enhanced sensitivity and resolution allow us to study disks in high-mass SFRs at kiloparsec-distance scales. One of those regions is NGC $6357$ ($d\approx 1690$ pc). This region hosts some of the most massive O2 and O3 stars known in the Galaxy \citep{Walborn2003}, along with an additional twenty O-type stars \citep{Russeil2010,Broos2013,Povich2013,Ramirez-Tannus2020}. The eXtreme Ultraviolet Environments (XUE) program (GO 1759, \citealt{Ramirez-Tannus2021}) targeted $12$ disks in NGC~6357 with the MIRI/MRS instrument. For those disks, the strength of the external FUV field ranges from $10^3$ to $10^6$ times the value in the solar neighborhood. \cite{Ramirez-Tannus2023} performs the initial characterization of one of those sources---the disk around the T Tauri star XUE 1. Using a slab modeling approach, they report the presence of \ce{CO, H2O, C2H2, HCN} and \ce{CO2}. Their retrieval approach suggests that the properties of those molecular tracers (emitting areas, column densities, and excitation temperatures) are indistinguishable from those observed in nearby, non-irradiated disks.            

While slab models are pivotal in the initial characterization of a disk, an in-depth understanding of its structure requires increasing modeling complexity. This next level of complexity materializes as thermochemical models, which are also crucial to inform future observational campaigns focusing on probing different physical conditions in the disk \citep{Kamp2015,Kamp2023}. We introduce the first thermochemical model, driven by JWST/MIRI and archival data, of a protoplanetary disk that is constantly exposed to an FUV field $10^5$ times stronger than the solar neighborhood value. Based on this model, we aim to constrain the physical structure of the XUE 1 disk.

Section \ref{sec:methods} introduces the data set used in this study as well as the modeling approach. Section \ref{sec:results} presents our constraints on the dust and gas distributions as well as the inferred gas temperature and molecular abundances of selected species which we obtain from the model that best explains the data. These results are contrasted to what would be expected if the disk were irradiated with a weaker FUV field. In Sect. \ref{sec:discussion}, we compare our results to those from slab models, elaborate on the key assumptions of our thermochemical  model, and relate our findings to the predictions from radiation-hydrodynamical models of photoevaporated disks. We present our conclusions in Sect. \ref{sec:conclusions}.

\begin{figure}
  \centering
  \begin{minipage}{0.45\textwidth}
    \centering
    \includegraphics[width=\textwidth]{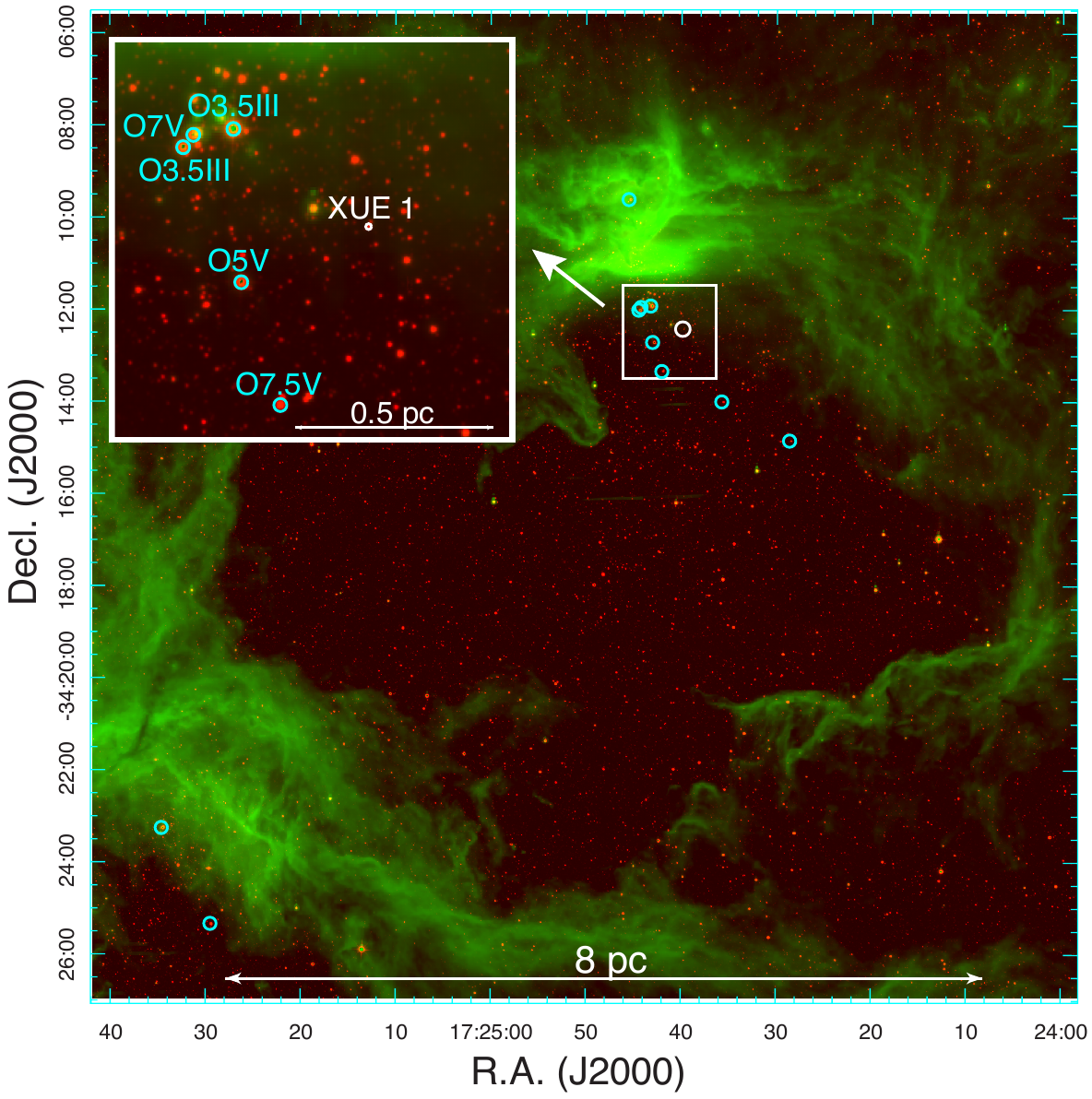}
  \end{minipage}
  \caption{Composite image of the CS 61 bubble in the NGC~6357 star-forming region, covering $20 \times 20$~arcmin$^2$ and combining UKIRT $K_s$-band (red) with Spitzer-IRAC 8.0 micron (green) data. The bubble hosts the rich Pismis 24 young stellar cluster. The $K_s$-band highlights the stellar component, including both members of the Pismis 24 cluster and unrelated field stars, while the 8.0 micron band reveals warm dust from the parental cloud, heated by intense radiation from massive stars. XUE 1 and known ionizing O-type stars are marked with white and cyan circles, respectively. Inset: a $2 \times 2$~arcmin$^2$ close-up of the XUE 1 neighborhood, with spectral types of the nearest O-type stars labeled.
}
  \label{fig:xue1_neighborhood}
\end{figure}

\section{Observations and Methods} 
\label{sec:methods}
\subsection{Stellar properties}
\label{sec:star_props}
XUE~1 is a stellar member of the rich young stellar cluster Pismis~24, situated within the massive star-forming region NGC~6357. The projected distances from XUE~1 to the most luminous nearby O-type stars range from 0.3 to 0.5 pc (Fig.~\ref{fig:xue1_neighborhood}).

Based on the {\it Chandra} X-ray and near-infrared (NIR) 2MASS and UKIRT $JHK_{s}$ data, combined with the theoretical predictions from the pre-main sequence (PMS) PARSEC 1.2S evolutionary model \citep{Bressan2012,Chen2014}, \citet{Getman2021,Getman2022} derive stellar properties for numerous young stellar members of over 40 nearby Galactic star-forming regions, including NGC~6357. These stellar properties include time-averaged X-ray luminosities, approximate stellar ages, source extinctions, effective temperatures, bolometric luminosities, and masses.

XUE~1 is the primary component (A1) in a binary \rt{stellar} system A1$+$A2, with a component separation of 0.2 arcseconds ($\sim 300$~au). \rt{The binary} remains unresolved in {\it Spitzer} mid-infrared (MIR), UKIRT, and VISTA NIR images, as well as {\it Chandra} X-ray images \citep{Povich2013,King2013,Broos2013,Townsely2019}. However, \rt{the XUE~1 stellar component} is resolved in the archived optical HST ACS and our JWST/MIRI MIR images.   

Assuming a total-to-selective extinction factor of $R_V = 3.3$ for NGC~6357 \citep{Massi2015, Russeil2017, Fouesneau2022} and using the extinction law from \cite{Gordon2023}, the binary system's  NIR-based extinction was inferred to be $A_V = 9.2$~mag, with an estimated age of $\sim 0.7$~Myr \citep{Getman2022}. These values are consistent with those of numerous nearby young stellar members of NGC~6357. Further assuming that the HST z-band magnitudes of the binary components are unaffected by accretion, and using the PARSEC 1.2S model along with the aforementioned source extinction and stellar age values, we derive stellar masses of 1.2 and 0.7~M$_{\odot}$ for the A1 (XUE~1) and A2 stellar components, respectively. The PARSEC-based stellar bolometric luminosity and effective temperature for XUE~1 are 3.9~$L_{\odot}$ and 4729~K, respectively.

The X-ray luminosity and column density measurements independently confirm the inferred mass of XUE~1. The X-ray luminosity of the whole binary system, $\log(L_X) = 30.4$~erg~s$^{-1}$, is dominated by XUE~1 \citep{Townsely2019}. This X-ray value is consistent with the stellar mass of $\sim 1$~M${_\odot}$ according to the well-known empirical PMS $L_X-M$ correlation \citep{Preibisch2005,Telleschi2007}. The gas column density of $\log(N_H) = 22.2$~cm$^{-2}$ is also in line with the dust visual extinction value, assuming the typical gas-to-dust ratio $N_{H}/A_V = 2 \times 10^{21}$~cm$^{-2}$ \citep{Zhu2017}. 

\subsection{Protoplanetary disk data}
\label{sec:observations}
\subsubsection{JWST/MIRI spectroscopy}
We summarize the main results from the JWST/MIRI/MRS observations of XUE 1 that will inform our modeling. For a detailed description of the data acquisition and analysis, refer to \cite{Ramirez-Tannus2023}. 

In this work, we use an updated reduction of the MIRI spectrum obtained with JWST Pipeline version 1.14.0. This new reduction achieves a signal-to-noise ratio of $110$ and a noise level of $\sigma=0.12$ mJy. We measure these values in the line-free interval $15.865<~\lambda<~15.952 \ \micron$. \rt{Our MIRI-MRS observations achieve an angular resolution at $7 \ \micron$ of $\sim 0.35$ arcsec \citep{Law2023}, which translates into $\sim 600$ au at the distance of NGC 6357.} We deredden the spectrum using a total-to-selective extinction factor $R_\mathrm{V}=3.3$, a visual extinction of $A_\mathrm{V}=9.2$ mag (Sect. \ref{sec:star_props}), and the \citealt{Gordon2023} extinction law. A visual inspection of the underlying continuum already confirms the presence of amorphous and crystalline silicates (we elaborate on the continuum characterization in Sect. \ref{sec:continuum-characterisation}). 

Line emission features in both MIRI channels 1 ($4.90 \leq \lambda \leq 7.65 \ \micron$) and 3 ($11.55 \leq \lambda \leq 17.98 \ \micron$) match their respective counterparts as seen in version 1.9.4 of the data reduction \citep{Ramirez-Tannus2023}. Thus, we confirm the presence of \ce{HCN}, \ce{H2O}, \ce{C2H2}, \ce{CO2}, and \ce{CO}. In addition to those features, our new reduction also displays a weak signal from \ce{OH} ($\sim 1$ mJy flux level) at $\sim \rt{16.0} \ \micron$ and $\sim 16.8 \ \micron$. \rt{These lines are pure rotational transitions of the ground electronic state and their quantum numbers are $(0,18.5,1,f)\rightarrow (0,17.5,1,f)$ and $(0,16.5,2,e)\rightarrow (0,15.5,2,e)$, respectively\footnote{\rt{Notation follows \cite{Brooke2016}. A quantum state is written as $(v,J,F,p)$, where $v$ indicates the vibrational level, $J$ is the total angular momentum quantum number, $F$ indicates the hyperfine state, and $p$ is the parity of the $\Lambda-$doubling state.}}.} 

\subsubsection{Archival photometry}
We use archival optical and NIR data from HST, VISTA and Spitzer. From these data, only HST is capable of resolving XUE 1 from the binary pair; accordingly, we adopt VISTA and Spitzer data as lower limits for the apparent magnitude of XUE 1. 

We correct for extinction in the same way as for the MIRI data. Zero-points and reference wavelengths for each filter are taken from the SVO Filter Profile Service \citep{Rodrigo2012,Rodrigo2020}. We summarize the photometry towards the source in Table \ref{table:photometry}. 

\renewcommand{\arraystretch}{1.25} 
\begin{table}
\fontsize{6}{6}\selectfont
\caption{Compilation of astrometric and photometric data towards XUE 1. Inequalities indicate lower (upper) limits on the magnitude (flux) from the unresolved A1 component.}
\label{tab:model_properties}
\centering
\begin{tabular}{llc}
\toprule 
R.A. & 17h 24m 40.100s \\
Dec & -34d 12m 25.36s  \\
Distance & 1690 pc  \\
\hline
\\
\multicolumn{3}{c}{Photometric properties}\\
\\
\hline
\textbf{Instrument} & \textbf{Magnitude} & \textbf{Extinction-corrected flux} \\
& & \textbf{($\mathrm{erg}\ \mathrm{s}^{-1} \ \mathrm{cm}^{-2} \ \mathrm{\AA}^{-1}$)} \\
\hline
HST/ACS - 550M              &  $V=22.7$        & $1.03\times 10^{-14}$\\
\hspace{1.15cm} - 850LP     & $z=17.9$         & $3.26\times 10^{-15}$ \\
VISTA/NIR                   & $J\geq 15.07$    & $\leq 3.31\times 10^{-15}$ \\
                            & $H\geq 12.99$    & $\leq 3.18\times 10^{-15}$ \\
                            & $K_\mathrm{s}\geq 12.80$    & $\leq 8.51\times 10^{-16}$ \\
Spitzer/IRAC.I1             & $[3.6 \ \micron]\geq 10.56$ & $\leq 5.96\times 10^{-16}$ \\ 
\hspace{0.9cm} IRAC.I2      & $[4.5 \ \micron]\geq 10.27$ & $\leq 2.81\times 10^{-16}$ \\
\hspace{0.9cm} IRAC.I3      & $[5.8 \ \micron]\geq 9.84$ & $\leq 1.56\times 10^{-16}$ \\ 
\hspace{0.9cm} IRAC.I4      & $[8.0 \ \micron]\geq 9.24$ & $\leq 8.19\times 10^{-17}$ \\
\hline
\end{tabular}
\label{table:photometry}
\end{table}

\subsection{Modeling}

\subsubsection{Continuum estimation and characterization}
\label{sec:continuum-characterisation}

\rt{In the following analysis of the dust population characterization, we first calculate the dust continuum emission level from the MIRI spectrum using the  \texttt{ctool} package \citep{Pontoppidan2024}, and then input this information into the Dust Continuum Kit \citep[\duck;][]{Kaeufer2024,Jang2024} package to derive the dust composition.}

We use the \texttt{ctool} package to estimate the underlying continuum in the $5 \leq \lambda \leq 28 \ \micron$ range of the MIRI spectrum. \texttt{ctool} estimates the continuum iteratively; it starts by median filtering the initial spectrum with a kernel size equal to \texttt{boxsize} wavelength channels. This filtering step produces a smoothed spectrum. The initial and smoothed spectra are compared to each other to generate a new continuum estimate: fluxes in the initial spectrum stronger than the smoothed flux are rejected and replaced by interpolating neighboring values \texttt{threshold} times weaker than the smoothed flux. These steps are repeated \texttt{niter} times, after which the remaining signal is convolved with a Savitzky-Golay filter. We run \texttt{ctool} using its default parameters: \texttt{niter=5}, \texttt{boxsize=95}, and \texttt{threshold=0.998}. As a pre-processing step, we first remove the broad molecular features using the slab models for \ce{C2H2}, \ce{HCN}, and \ce{CO2} found by \cite{Ramirez-Tannus2023}.

We perform a retrieval analysis to identify which dust species are shaping the continuum in the mid-IR; this will inform the thermochemical code about an appropriate set of dust opacities to use when solving the continuum radiative transfer. However, this work does not aim at a detailed study of the mineralogy in XUE 1.  

We use the Dust Continuum Kit (\duck) package. \duck \ retrieves the grain composition in two steps. First, \duck \ simulates the continuum emission as a linear combination of four flux terms: a star, an optically thick inner rim, an optically thick midplane, and an optically thin surface. Particularly, the flux from the optically thin surface is modeled as the sum of individual terms over the number of dust species considered. Each individual term is proportional to the radial integral of a Black-Body function weighted by the optical depth of the respective dust species\footnote{We use dust opacities computed with the Distribution of Hollow Spheres formalism \citep{Min2005}.}. In a second step, \duck \ generates models for the total flux using Multimodal Nested Sampling and determines the best fit parameters, given an observation, by evaluating a likelihood function. 

\begin{figure}
  \centering
  \begin{minipage}{0.45\textwidth}
    \centering
    \includegraphics[width=\textwidth]{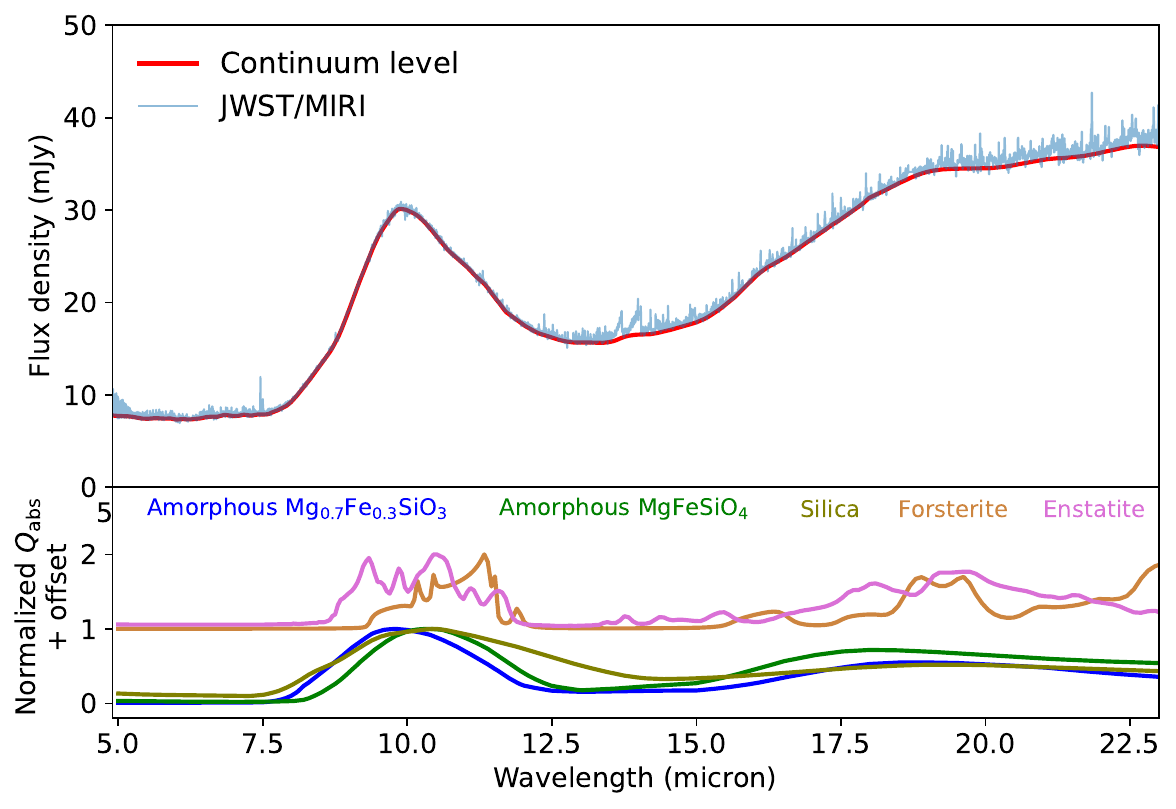}
  \end{minipage}
  \caption{\rt{Characterization of the optically thin dust in the XUE 1 disk. \textit{Top:} the blue curve is the total (dust+lines) JWST/MIRI spectrum and the red curve is the continuum level estimated with \texttt{ctool}. \textit{Bottom:} Absorption efficiency curves. The curves are for $1\ \micron$ pure grains whose composition is color-coded.}}
  \label{fig:xue1_dust}
\end{figure}

\duck \ finds evidence\footnote{Evidence for the presence of a species means that \duck \ retrieves a total mass fraction higher than $1 \%$ for that species.} for both amorphous and crystalline silicates. In the amorphous category, we find silica ($\silica$), and amorphous silicates with $\mathrm{SiO}_3$ stoichiometry ($\pyroxene$) and $\mathrm{SiO}_4$ stoichiometry ($\olivine$). In the crystalline category, we find forsterite ($\forsterite$) and crystalline $\enstatite$ (that we refer hereafter as enstatite). We do not find evidence for iron-rich minerals such as fayalite ($\fayalite$) and troilite (FeS). The non-detection of troilite is expected given the absence of strong solid state resonances within the MIRI range. \rt{The estimated continuum level and the absorption efficiency curves of the retrieved species are shown in Fig. \ref{fig:xue1_dust}.}

\subsubsection{Thermochemical modeling}
\label{sec:modelling}
We use the \prodimo \ code \citep{Woitke2009,Kamp2010,Thi2011,Woitke2016,Rab2018,Woitke2024} to perform thermochemical modeling of the XUE 1 disk. \prodimo \ is a radiation thermochemical code that self-consistently solves for the continuum radiative transfer, the gas heating-cooling balance and the chemistry in a protoplanetary disk. The code assumes a two-dimensional geometry implemented on a cylindrical grid ($r,z$), where $r$ is the radial coordinate measured from the inner rim, and $z$ is the vertical coordinate measured from the disk midplane. Simulations are carried out with \prodimo \ version \texttt{V.3.0.0}, revision \texttt{9f058bf9} \citep{Woitke2024}. This is an improved and expanded version of the code that is capable to fit JWST spectra. Some of the new features include: an improved escape probability algorithm to calculate line fluxes and emitting areas \citep{Woitke2024}; spectroscopic data for LTE calculations taken from HITRAN 2020 \citep{Gordon2022} database, and the addition of new data for selected hydrocarbons \citep{Arabhavi2024}; an improved treatment for the calculation of UV photo-rates and molecular shielding factors \citep{Woitke2024}.  

In light of the scarcity of data, we built on two working assumptions and a series of initial conditions. Our assumptions are based on either well-established observational results or theoretical grounds. 

\textit{\textbf{Assumption 1:} XUE 1 is a primordial disk.} We assume the disk has no substructures, and that the dust and gas distributions follow a continuous, exponentially decaying profile in the radial direction and a Gaussian profile in the vertical direction. The gas and dust surface densities can therefore be written as

\begin{equation}
\label{eq:densities}
    \Sigma (r)=\Sigma_0 \bigg(\frac{\Rtap}{r}\bigg)^\epsilon \exp \bigg[-\Big(\frac{r}{\Rtap}\Big)^{2-\gamma}\bigg],
\end{equation}

\noindent where the constant $\Sigma_0$ scales with the disk mass and $\Rtap$ is the tapering radius; i.e., the distance marking the transition from a linear to an exponential decay of the surface density profile. We assume a self-similar constraint on the exponents, setting $\epsilon=\gamma=1$ \citep{Lynden-Bell1974,Hughes2008}. Vertical stratification of dust grains follows \cite{Dubrulle1995} with an $\alpha$ \citep{Shakura1973} parameter of $10^{-2}$.

\textit{\textbf{Assumption 2:} XUE 1 is constantly exposed to an interstellar FUV field characterized by $\Go=10^5$.} Hereafter, we adopt $\Go$ as the ratio of the local FUV field to that in the solar neighborhood from \cite{Draine1978}:  

\begin{equation}
\label{eq:Gknot}
\Go \equiv \int_{91.2 \ \mathrm{nm}}^{205 \ \mathrm{nm}} \lambda u_\lambda \mathrm{d} \lambda \Bigg / \int_{91.2 \  \mathrm{nm}}^{205 \ \mathrm{nm}} \lambda u_{\lambda,\mathrm{Draine}} \mathrm{d} \lambda,
\end{equation}

\noindent where $u_\lambda$ is the specific energy density. We assume here that the three-dimensional separation of XUE 1 from the neighboring O stars is not much larger than the projected separation $\sim 0.5$ pc (see Fig. \ref{fig:xue1_neighborhood}). 

Next, we define the physical parameters and boundary conditions of a fiducial model for the XUE 1 disk. This fiducial model shares many similarities with the DIANA model for a standard T Tauri disk, introduced in \cite{Woitke2016} (see their Table 3). However, our fiducial disk differs in the following key aspects: the strength of the external FUV irradiation, in response to our second working assumption; the grain composition, in response to our characterization of the continuum (Sect. \ref{sec:continuum-characterisation}); and the disk mass, which is an observationally motivated choice as explained below. 

\textbf{Disk mass:} Since there is currently no direct or indirect tracer for the disk mass, we assume a fiducial dust mass of $\Md = 10^{-6} \ M_\odot$. The fiducial gas mass is obtained by rescaling $\Md$ by a canonical interstellar medium gas-to-dust ratio of $100$ (e.g., \citealt{Williams2021}), i.e., $\Mg=10^{-4} \ M_\odot$. We note that our choice for the dust mass is within the range $0.1-1.0 \ M_\oplus$, and is consistent with 60\% of the stellar disks observed in Lupus and Chamaeleon \citep{Tychoniec2018}. Similar trends have been observed for disks in Orion, although limited by lower sensitivities \citep{vanTerwisga2022}.        

\textbf{Disk size:} We assume a fiducial tapering radius $\Rtap=100$ au, following the DIANA standard T Tauri disk \citep{Woitke2016}. While this $\Rtap$ is consistent with the disk sizes observed in Ophiuchus  and Taurus \citep{Tripathi2017}, it is not consistent with the substantially lower distribution of disk sizes observed in Orion \citep{Eisner2018}. In Sects. \ref{sec:dust-distribution} and \ref{sec:gas-distribution}, we present model-based evidence that rules out our initial assumption of $\Rtap=100$ au, supporting a more compact configuration for the XUE 1 disk. Following \cite{Gullbring1998}, the inner disk radius is assumed to be within the corotation radius, at $R_\mathrm{in}\equiv 5 R_* = 0.07$ au, where $R_*$ is the stellar radius.   

\textbf{Grain properties:} We implement an MRN size distribution \citep{Mathis1977} between a minimum and maximum grain sizes of $0.05 \ \micron$ and $3$ mm, respectively. The MRN distribution is also characterized by a slope $p$---this parameter plays an important role in determining the shape of the SED at long wavelengths in the mid-IR. The MNR distribution has a slope $p=3.5$ as measured in the interstellar medium. However, detailed studies of spatially resolved disks seem to favor a range of values for the slope that go above and below $3.5$ (e.g., \citealt{Macias2021,Gauidi2022,Doi2023}). In particular, values higher than $p=3.5$ enhance the contribution of smaller grains to the total opacity and lead to a stronger $10\micron$ emission feature. For the fiducial model we adopt $p=3.9$, which is similar to the value found with \prodimo \ models for PDS 70 \citep{Portilla-Revelo2022} and EX-Lupi \citep{Woitke2024}. The size distribution is discretized in $100$ size bins. Grains are made of a mixture of carbonaceous ($15\%$ volume fraction) and silicate material ($60\%$), with $25\%$ of the volume of the aggregate assumed as vacuum. We analyze a small grid of models where only the continuum radiative transfer is solved and the SEDs are qualitatively compared to the MIRI continuum. This analysis suggests a crystallinity value of less than $5\%$ by volume. Additionally, for the amorphous silicates found in Sect. \ref{sec:continuum-characterisation}, the abundance is suggested to be in a $10:10:1$ ratio by volume. Table \ref{tab:model_properties} lists the volume fractions of silicates distributed among the elemental species retrieved in Sect. \ref{sec:continuum-characterisation}. We assume that the composition inferred from the mid-IR emission applies everywhere in the disk. Although this assumption seems valid given the scarcity of data, we note that certain properties might vary across the disk. For example, the value of the crystallinity could be spatially dependent if localized regions of efficient grain growth existed in the disk (e.g., \citealt{vanBoekel2005}).        

\textbf{Disk geometry:} The fiducial model is a flared disk in radiative equilibrium. The gas scale height is parameterized as $H(r)=H_0(r/r_0)^\beta$. Values for the scale height at the reference distance ($H_0$ and $r_0$), as well as for the flaring exponent ($\beta$), are listed in Table \ref{tab:model_properties}. Values for the reference scale height and the flaring exponent are slightly fine-tuned to enhance the overall quality of the fit. Nevertheless, the value for these parameters listed in Table \ref{tab:model_properties} are still within the ranges that explain the observed SEDs from other disks studied the DIANA project \citep{Woitke2019,Kaeufer2023}. Notably, a larger value for the reference scale height is theoretically expected for an irradiated disk since in hydrostatic equilibrium $H\equiv c_\mathrm{s}/\Omega \sim \sqrt{T_\mathrm{gas}}$, where $c_\mathrm{s}$ is the sound speed and $\Omega$ is the keplerian frequency. Finally, the inclination of the disk respect to the plane of the sky is set to $60^\circ$ following \cite{Ramirez-Tannus2023}. 

\rt{\textbf{Interstellar UV radiation field:} The UV component of the background interstellar radiation field ($I_\nu^{\mathrm{ISM,UV}}$) is assumed isotropic. It is modeled as a diluted black body at a temperature $20000$ K \citep{Woitke2016},}

\begin{equation}
    I_\nu^{\mathrm{ISM,UV}} = \chi \cdot 1.71 W_\mathrm{dil} \cdot B_\nu(20000 \mathrm{\ K}),
\end{equation}

\noindent \rt{with $W_\mathrm{dil}=9.85357 \times 10^{-17}$, a dilution factor such that $\chi=1$ (see Eq. \ref{eq:Gknot}) corresponds to the UV radiation field in the solar neighborhood. The DIANA standard model has $\chi=1$ and our fiducial model has $\chi=10^5$.}

\prodimo \ first solves the continuum radiative transfer and then iterates over the gas heating-cooling balance and the chemistry. Iterations are needed since the gas temperature and molecular abundances depend on each other. \prodimo \ computes the gas temperature by balancing several heating and cooling mechanisms; this step requires knowing the level populations of the gas species included in the model. Level populations are computed assuming either LTE or non-LTE conditions, depending on the availability of collisional data \rt{The levels of \ce{C2H2, HCN} and \ce{CO2} are all populated according to LTE. A non-LTE treatment is followed for \ce{H2O} (collision data are taken from the LAMDA database, see \citealt{Schoier2005,vanderTak2020} and references therein); pure rotational \ce{OH} lines \citep{Offer1994,Rahmann1999,Tabone2021}; and for \ce{CO}, for which the custom molecular model by \cite{Thi2013} is used. Spectroscopic data are mostly taken from the HITRAN \citep{Gordon2022} and LAMDA databases.}. An escape probability formalism determines the excitation and de-excitation rates. For the chemistry, we use the large chemical network from \cite{Kamp2017} with $236$ species and $3046$ chemical reactions. Creation and destruction processes and their respective rate coefficients are mostly taken from the UMIST database \citep{McElroy2013}. For a description of a few additional processes that are not in UMIST---such as freeze-out and desorption ice chemistry, and \ce{H2} formation via dust catalysis---we refer the reader to \cite{Woitke2024} and references therein for details. Finally, ray-tracing and line-escape probability algorithms are used to generate synthetic spectral energy distributions and emission line spectra. 

\section{Results} 
\label{sec:results}

\subsection{Physical structure of the XUE 1 disk: Dust and gas distribution}

\subsubsection{Dust distribution}
\label{sec:dust-distribution}
\begin{figure*}
    \centering
    \includegraphics[width=\textwidth]{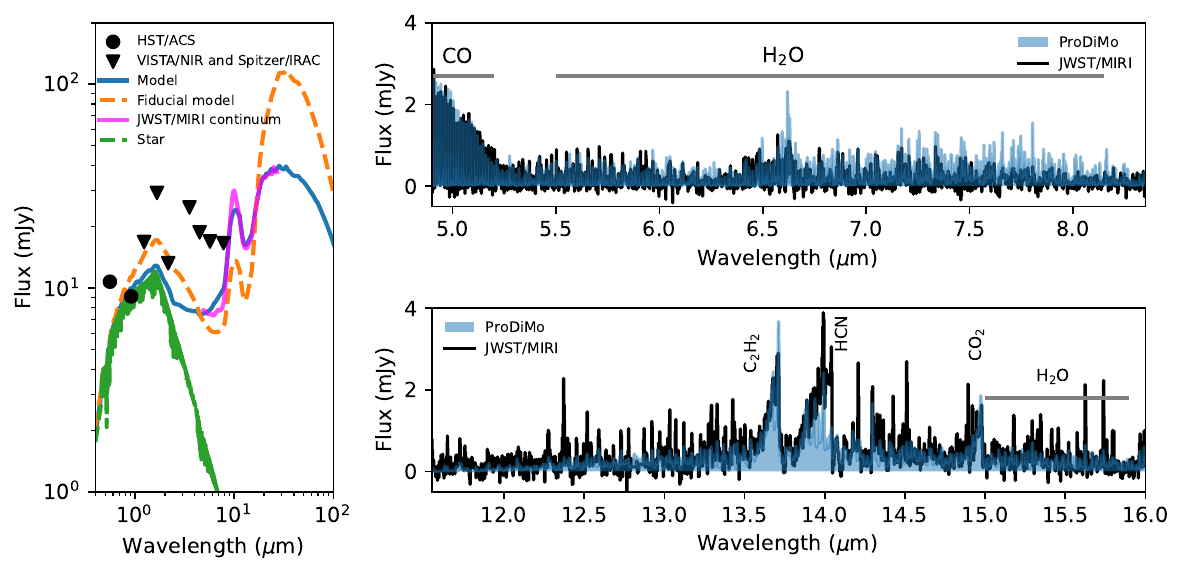}
    \caption{Synthetic predictions from the XUE 1 model against the observations. Left: Photometry towards XUE 1---dots indicate the HST/ACS photometry and downward arrows indicate upper limits from VISTA and Spitzer. The magenta line is the JWST/MIRI continuum. The green is the stellar spectrum, and the blue and orange lines indicate the spectral energy distributions of the best fit and fiducial disk models, respectively. Right: continuum-subtracted spectra for MIRI channels 1 (top) and 3 (bottom)---the black curves are the MIRI data and the areas filled in blue are the \prodimo \ spectra. Synthetic spectra are convolved using a Gaussian kernel with $R=3500$ for channel 1, and $R=2500$ for channel 3.}
    \label{fig:synthetic-observables}
\end{figure*}

The model for the fiducial disk performs poorly in explaining the observations. Notably, the response to an enhanced FUV irradiation is evident on the continuum: the synthetic SED displays very strong mid- and far-infrared emission that is not consistent with the observations (see left panel in Fig. \ref{fig:synthetic-observables}). To quantify this discrepancy, we use the spectral index 

\begin{equation}
\nspectral \equiv \frac{\log(\lambda_{25}F_{25}) - \log(\lambda_{13}F_{13})}{\log(\lambda_{25}) - \log(\lambda_{13})},
\end{equation}

\noindent where $F_\lambda$ is the specific flux and the wavelength $\lambda$ is in microns. The spectral index predicted by the fiducial model is $1.80$ times higher than the observed one $\nspectral^\mathrm{obs}=5.53$. This behavior is a consequence of Kirchhoff's law---the mass absorption coefficient of the dust distribution implies a high efficiency of absorption of UV photons, which in turn implies an efficient re-emission at infrared wavelengths.

For a flared disk in radiative equilibrium, the emission in the $10\lesssim \lambda \lesssim 100\ \micron$ range is dominated by an optically thin surface of  dust \citep{Chiang1997}. Thus, the flux scales with the disk's solid angle as seen from the observer. Accordingly, we reduce the disk's size by cutting the tapering radius down from $100$ au to $10$ au, while keeping the dust mass constant. This change renders the fiducial disk into a compact disk, for which the emitting area decreases and the density in the inner regions increases. The synthetic SED from the compact disk has a spectral index that is only $1.01$ times the observed one. We note that our choice of $\Rtap=10$ au comes from comparing only the observed to synthetic spectral indexes. A thorough study on the effect of the tapering radius on the synthetic line emission, as well as the impact of Polycyclic Aromatic Hydrocarbons (PAHs), will be presented in S. Hernández et al. (in prep). We also disregard the tidal effects from the A2 component (the secondary member of the binary pair). The compactness of the system and the relatively low mass of the secondary (Sect. \ref{sec:star_props}) make it unlikely that tidal effects could truncate the disk below $10$ au \citep{Panic2021}. Similarly, we neglect any heating contribution due to the radiation field set by the companion.            

\begin{figure}
  \centering
  \begin{minipage}{0.45\textwidth}
    \centering
    \includegraphics[width=\textwidth]{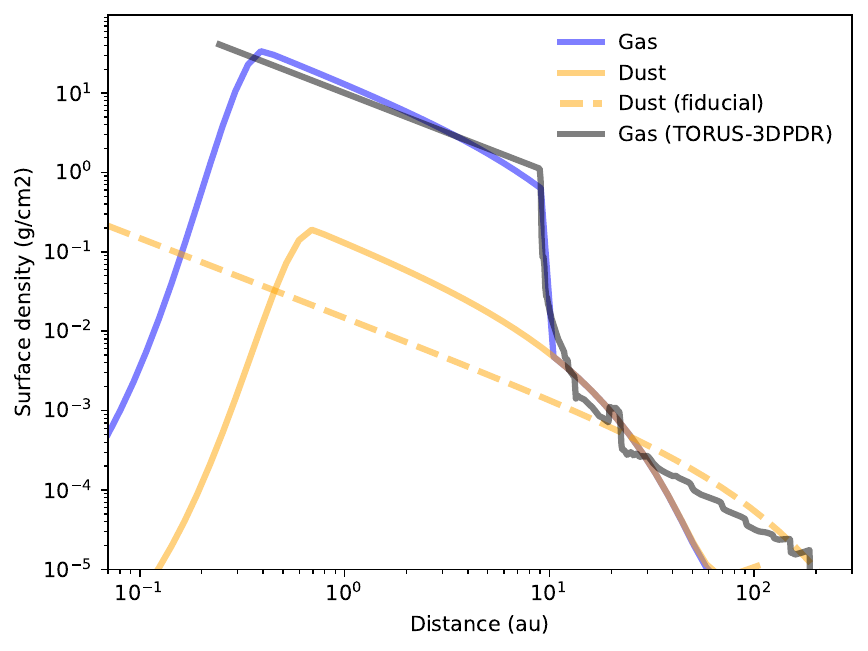}
  \end{minipage}
  \caption{Surface density profiles of dust and gas as a function of semi-major axis. Solid curves indicate the solution from the best representative model for XUE 1. The dashed line indicates the dust surface density of the fiducial disk. Black line is the steady-state solution of the radiation-hydrodynamical model discussed in Sect. \ref{sec:hydro}.}
  \label{fig:dust-gas-distribution}
\end{figure}

The continuum emission at near- and short mid-IR wavelengths is sensitive both to the geometry and to the density distribution near the disk's inner rim \citep{Dullemond2010}. These two properties determine the radial optical depth in the inner disk. Equation (\ref{eq:densities}) implies a step function-like transition of the density at the inner radius. Since this is clearly unrealistic, we further modify the initial dust density profile to allow for a gradual build up of solid material from the inner radius outwards. This is modeled as a depletion of the initial profile by a Gaussian function\footnote{The choice of a Gaussian function is arbitrary. However, this is a simple way to simulate smooth changes in density that are also axisymmetric. This approach has also been applied to describe the dust distribution near gaps and cavities in disks (e.g., \citealt{Keppler2018}).} centered at $r=0.6$ au with a standard deviation of $0.15$ au. The center of the Gaussian and its standard deviation come from a small grid of models aimed at explaining the observed NIR photometry and the MIRI continuum around the $10 \ \micron$ feature. This approach is similar in nature to the method introduced in \cite{Woitke2024} to simulate the gradual build up of dust near the inner rim. However, our approach can treat the dust and gas density distributions independently from each other. Figure \ref{fig:dust-gas-distribution} shows the dust surface density profiles for both the fiducial and best fit models. We highlight that the dust surface density at $0.13$ au---where the equilibrium temperature of a grain reaches the silicate sublimation threshold of $1500$ K---is four orders of magnitude lower compared to the unperturbed profile of the fiducial model. Qualitatively, this is in accordance with what is expected in the presence of dust sublimation driven by stellar radiation. 


The synthetic SED from the best fit model is compared to the observation in the left panel of Fig \ref{fig:synthetic-observables}. Integrating the final surface density profile over the radial and azimuthal coordinates, we find a dust mass of $0.3 \ \Mearth$. Thus, the dust mass of the best representative model is only $10\%$ lower than the initial value, due to the enforced dust depletion near the inner rim. 

\subsubsection{Gas distribution}
\label{sec:gas-distribution}
We generate a continuum-subtracted spectrum starting from the disk model that already explains the continuum (Sect. \ref{sec:dust-distribution}). We first focus on the wavelength range covered by MIRI-channel 1. As observed in nearby disks around T Tauri stars, this spectral window shows prominent emission from \ce{H2O and CO} \citep{Kospal2023,Perotti2023,Gasman2023,Henning2024}. This is also the case for the XUE 1 disk \citep{Ramirez-Tannus2023}.

\begin{figure}
  \centering
  \begin{minipage}{0.45\textwidth}
    \centering
    \includegraphics[width=\textwidth]{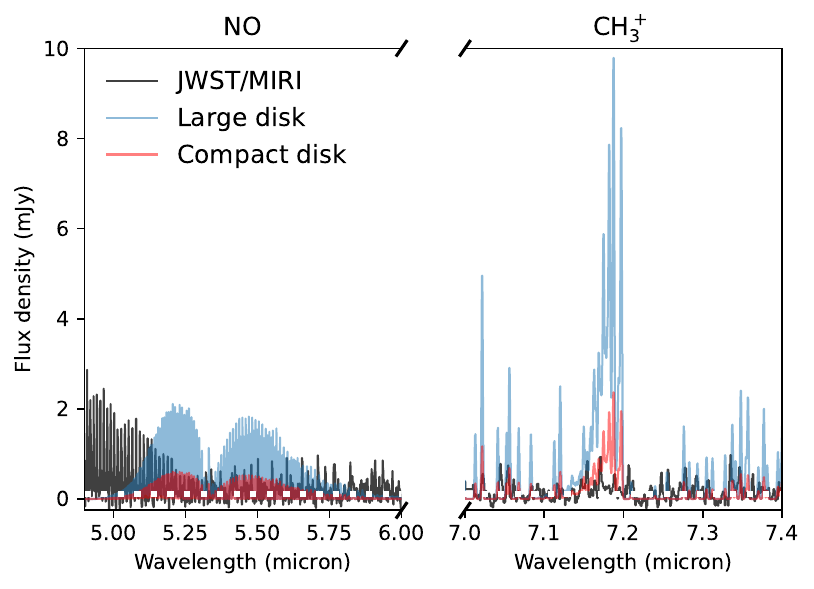}
  \end{minipage}
  \hfill
  \begin{minipage}{0.45\textwidth}
    \centering
    \includegraphics[width=\textwidth]{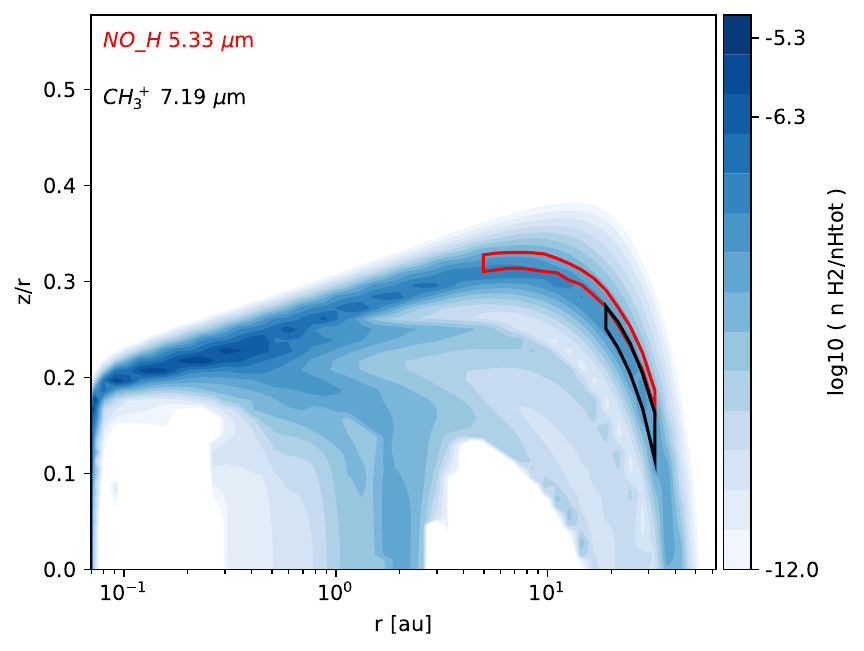}
  \end{minipage}
  \caption{Synthetic predictions for \rt{\ce{NO} and \ce{CH3+}}. Top: continuum-subtracted spectra---blue and red lines are from the fiducial and compact disks, respectively. The black line is the MIRI observation and the white dots indicate the noise level. The observed features below $5.2 \ \micron$ correspond to the P-branch of the \ce{CO} rovibrational emission. Bottom: emitting areas of the \ce{NO} (red box) \rt{and \ce{CH3+} (black box)} emission \rt{retrieved from the compact disk}. Color map shows the gas-phase abundance of \ce{NO} relative to the total hydrogen nuclei.} 
  \label{fig:NO-emission}
\end{figure}

The synthetic spectrum showed an emission feature around $5.4 \ \micron$ that contrasts with the data: the P-, Q-, and R-branches of nitrogen monoxide (\ce{NO}) ro-vibrational emission (Fig. \ref{fig:NO-emission}-top). For the compact disk, the strength of this feature is more than twice the noise level; for the fiducial disk, the strength is almost ten times the noise level. Further inspection of the models revealed that this emission originates in the outer disk---beyond $15$ au for the compact disk (Fig. \ref{fig:NO-emission}-bottom), and beyond $30$ au for the fiducial disk.

\ce{NO} is a diatomic molecule whose formation and destruction pathways, for large $r$ and $z$ values, proceed respectively via \citep{Schwarz2014}: 

\begin{equation}
\label{eq:NO-chem}
\begin{split}
    \ce{N + OH &-> NO + H,}\\
    \ce{N + NO &-> N2 + O},
\end{split}    
\end{equation}

\noindent and thus, a high abundance of \ce{NO} should be accompanied with a high abundance of \ce{OH}.  In dense environments, formation of \ce{OH} via dissociative recombination of \ce{H3O+} is favored by a large abundance of free electrons\footnote{\rt{For the fiducial disk externally irradiated with $\chi=10^5$, the electron abundance averaged over the emitting region of \ce{NO} is $\mathcal{O} (10^3 \ \mathrm{cm}^{-3})$. For the non-irradiated fiducial disk ($\chi=1$), the averaged electron abundance is only $\mathcal{O} (10^2 \ \mathrm{cm}^{-3})$.}}. In an irradiated disk, FUV photoionization of abundant species such as carbon and sulfur, increases the number of free electrons in the disk. \rt{Likewise, gas-phase oxygen chemistry driven by warm \ce{H2} leads to \ce{OH} formation via the reaction \ce{H2 + O \rightarrow OH + H}, a process recently observed in the environment of an irradiated disk in Orion \citep{Zannese2024}.} These arguments naturally explain the high abundance of \ce{NO} both in the fiducial and compact disks, which are both strongly irradiated. 


While a lower abundance of free electrons (i.e., a weaker external FUV field) can reduce that of \ce{NO} in the outer disk, a lower gas surface density beyond $10$ au can also have the same effect. We explore the latter option and defer the former for the discussion. We deplete the gas mass beyond $10$ au by factors of $10$ and $100$ with respect to the nominal values in the fiducial model---the latter depletion factor implies a gas-to-dust ratio of $1.0$ in the outer disk. This gas-depleted and compact disk shows no \ce{NO} emission above the noise level, in accordance to the observation. 

One caveat is that our model is limited by the lack of collisional data for \ce{NO}, and therefore we assume Local Thermodynamic Equilibrium (LTE) for its level populations. However, since the \ce{NO} emission comes from the outer disk (Fig. \ref{fig:NO-emission}), LTE might not accurately reflect the excitation conditions in this low-density area. \rt{We estimate a volume-weighted average abundance of total hydrogen nuclei within the \ce{NO} emitting region of $\langle n (\mathrm{H_{tot}}) \rangle = 3.0\times 10^7 \ \mathrm{cm}^{-3}$}. A non-LTE approach could result in lower upper state population densities, as seen with other molecules like \ce{CO} \citep{Thi2013}. Thus, non-LTE effects might reduce the need to remove gas from the outer disk. A detailed study of \ce{NO}'s spectral signature requires rovibrational collisional data, which is currently unavailable in the literature. Determining these collisional coefficients is well beyond the scope of this work.

\rt{Motivated by the recent detection of \ce{CH3+} in the wind of an irradiated disk in Orion \citep{Berne2023}, we generate a synthetic spectra for this molecule in MIRI/channel 1\footnote{\rt{The reader is cautioned that all results related to \ce{CH3+} were obtained using a beta version of \prodimo, rather than the current stable version (v3.0.0) used in the rest of the paper. While we acknowledge the potential limitations this may introduce for reproducibility, we believe the results are robust enough to be included in this manuscript.}}. Similar to \ce{NO}, our model predicts a very strong emission feature around $\sim 7.2 \ \micron$, which contrasts sharply with the observation. The bottom panel in Fig. \ref{fig:NO-emission} shows that this emission also originates from the outer disk. Since we de not detect \ce{CH3+} in the XUE 1 disk, we interpret this result as further evidence for the need of gas depletion in the outskirts of the disk.} 

Finally, starting now from the compact and outer gas-depleted disk, we generate a synthetic spectrum that covers MIRI-channel 3, where strong emission from \ce{C2H2, HCN and CO2} is observed in XUE 1 \citep{Ramirez-Tannus2023}. The synthetic spectra already agree with the observation within a factor of two for each of those molecules.

The largest discrepancy occurs in channel 1, where \prodimo \ predicts water lines that are up to a factor of $4$ stronger than the observation. The model indicates that the emitting region of those lines is restricted to the inner $1$ au (see Sect. \ref{sec:emitting-regions} for a discussion on line emitting regions). We take the \ce{o-H2O} $\lambda \approx 6.62 \ \micron$ line as a proxy: this line has an emitting region that extends vertically from $z/r=0.16$ down to $z/r=0.045$, and radially from $r=0.07$ au out to $r=0.4$ au. In order to reduce the \ce{o-H2O} $\lambda \approx 6.62 \ \micron$  line strength in the model, we modify the inner disk gas distribution using a Gaussian depletion function centered at $r=0.4$ au. Exploration of a coarse grid of models suggests a value for the standard deviation of $0.07$ au. We determine this value for the standard deviation by visually comparing the synthetic and observed line fluxes in channels 1 and 3. A stronger depletion would further improve the fit to the water features (by making them weaker), but it would underfit the \ce{CO} flux in the P-branch. The final gas density profile is shown in Fig. \ref{fig:dust-gas-distribution}.

Table \ref{tab:model_properties} summarizes the parameters of the best fit model. The continuum-subtracted synthetic spectra are compared to the observations in the right panel of Fig. \ref{fig:synthetic-observables}. Integrating the gas surface density yields a total gas mass of $\sim 6\times 10^{-5} \ M_\odot$ for the XUE 1 disk.

\begin{deluxetable}{ll}
\tabletypesize{\normalsize}
\tablewidth{\columnwidth} 
\tablecaption{Parameters of the best representative model for the XUE 1 disk}

\tablehead{Parameter & Value} 

\startdata 
Mass & $1.2 \ \mathrm{M}_\odot$\\
Bolometric luminosity ($L_\mathrm{bol})$ & $3.9 \ (L_\odot)$\\
Effective temperature & $4729\, \mathrm{K}$\\
$L_\mathrm{FUV}/L_\mathrm{bol}$\tablenotemark{a} & $0.04$ \\
X-ray luminosity  ($L_\mathrm{X}$) & $2.5 \times 10^{30} \ \mathrm{erg} \ \mathrm{s}^{-1}$ \\
X-ray emission temperature & $3\times 10^7$ K\\
Mass accretion rate\tablenotemark{b} & $1.76 \times 10^{-9} \ M_\odot \ \mathrm{yr}^{-1}$\\
Relative external FUV field ($\Go$) & $10^5$ \\
\hline
Minimum grain size & $0.05 \ \micron$ \\
Maximum grain size & $3000 \ \micron$ \\
Grain size power index & 3.9 \\
Porosity & $25 \%$ \\
Amorph. carbon (by volume) & $15 \%$ \\
Amorph. pyroxene ($\pyroxene$) & $26.2 \%$ \\
Amorph. olivine ($\olivine$) & $26.2 \%$ \\
Amorph. Silica ($\silica$) & $2.6 \%$ \\
Forsterite ($\forsterite$) & $2.5 \%$ \\
Enstatite ($\enstatite$) & $2.5 \%$ \\
Turbulent settling parameter ($\alpha$) & $10^{-2}$ \\
\rt{Gas phase} carbon-to-oxygen ratio & $0.457$\\
\hline
Disk gas mass & $6.1 \times 10^{-5} \ M_\odot$ \\
Disk dust mass & $0.3 \ M_\oplus$ \\
Inner radius & 0.07 au \\
Tapering radius ($\Rtap$) & 10 au \\
Inclination & $60^\circ$  \\
Flaring exponent ($\beta$) & 1.30 \\
Reference radial distance ($r_0$) & 10 au \\
Scale height at $r_0$ ($H_0$) & 1 au \\
\enddata
\tablenotetext{a}{\small We treat the stellar FUV luminosity as a free parameter. A $L_\mathrm{FUV}$ value that is due to accretion only cannot explain the strength of the \ce{CO} emission. This suggests that chromospheric activity plays a major role in setting the FUV luminosity of XUE 1.}
 \tablenotetext{b}{\small Average value observed in Taurus by \cite{Lin2023}}
\label{tab:model_properties}
\end{deluxetable}

\subsection{Physical structure of the XUE 1 disk: Gas temperature and molecular abundance}
\label{sec:temperatures-and-abundances}
\begin{figure*}[h!]  
  
  \centering
  
  \begin{minipage}{0.3\textwidth}
    \centering
    \includegraphics[width=\textwidth]{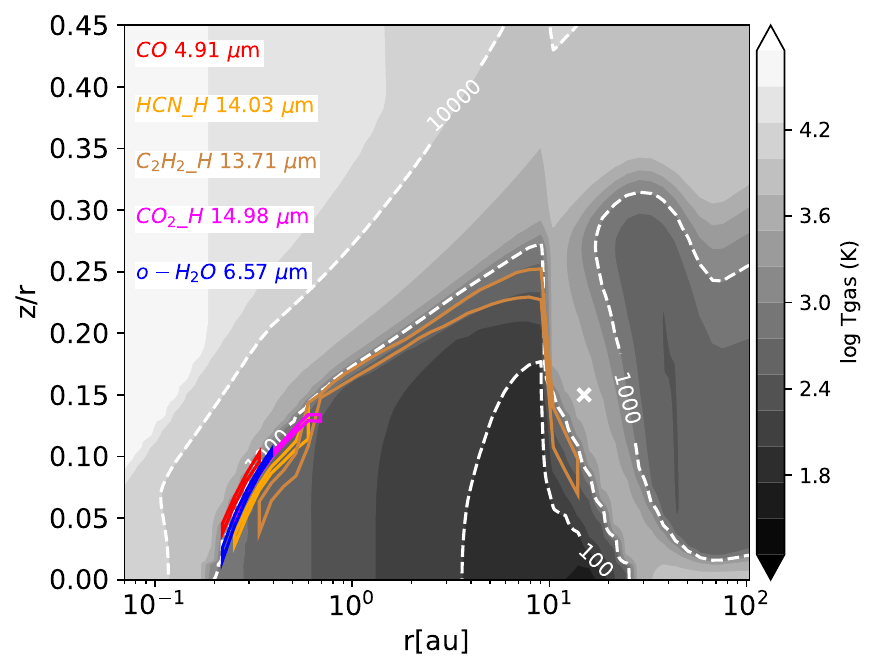}
  \end{minipage}
  \hfill
  \begin{minipage}{0.3\textwidth}
    \centering
    \includegraphics[width=\textwidth]{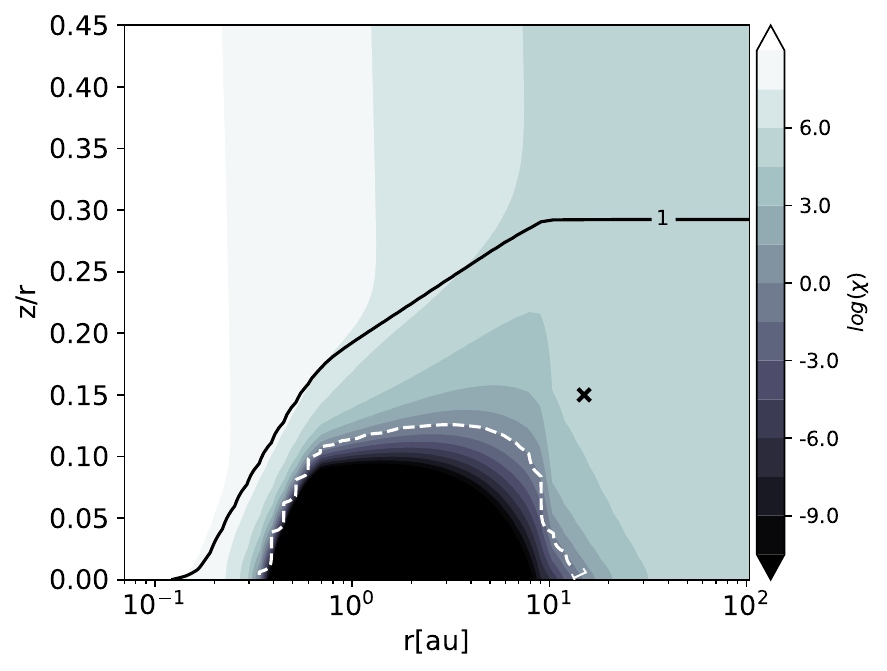}
  \end{minipage}
  \hfill
  \begin{minipage}{0.3\textwidth}
    \centering
    \includegraphics[width=\textwidth]{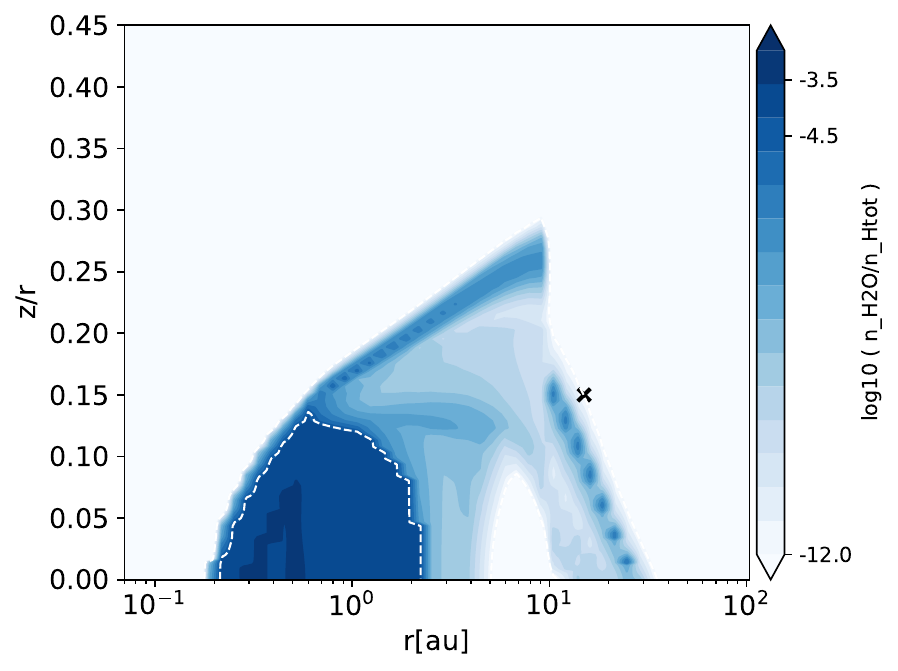}
  \end{minipage}

  \vfill

  \begin{minipage}{0.3\textwidth}
    \centering
    \includegraphics[width=\textwidth]{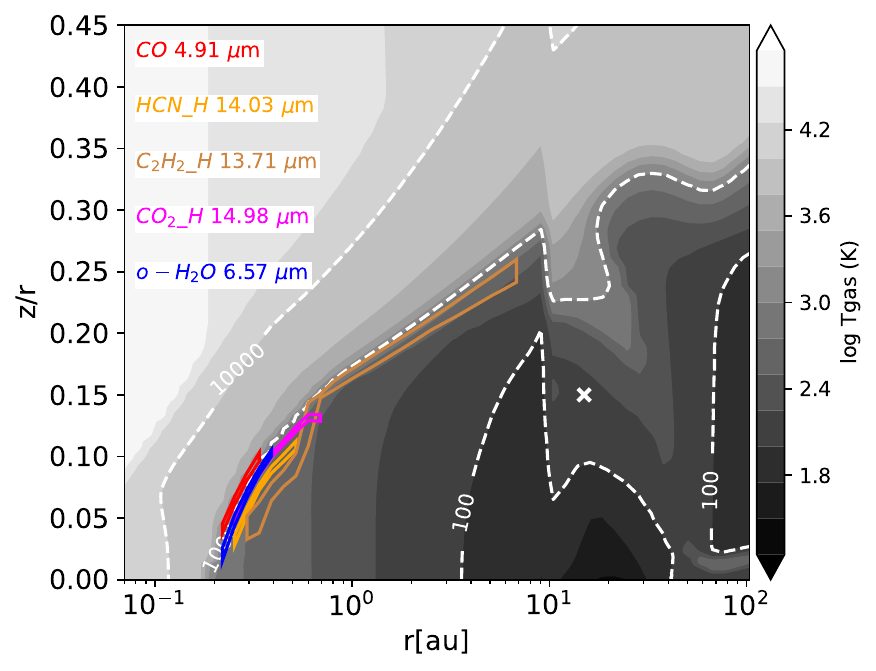}
  \end{minipage}
  \hfill
  \begin{minipage}{0.3\textwidth}
    \centering
    \includegraphics[width=\textwidth]{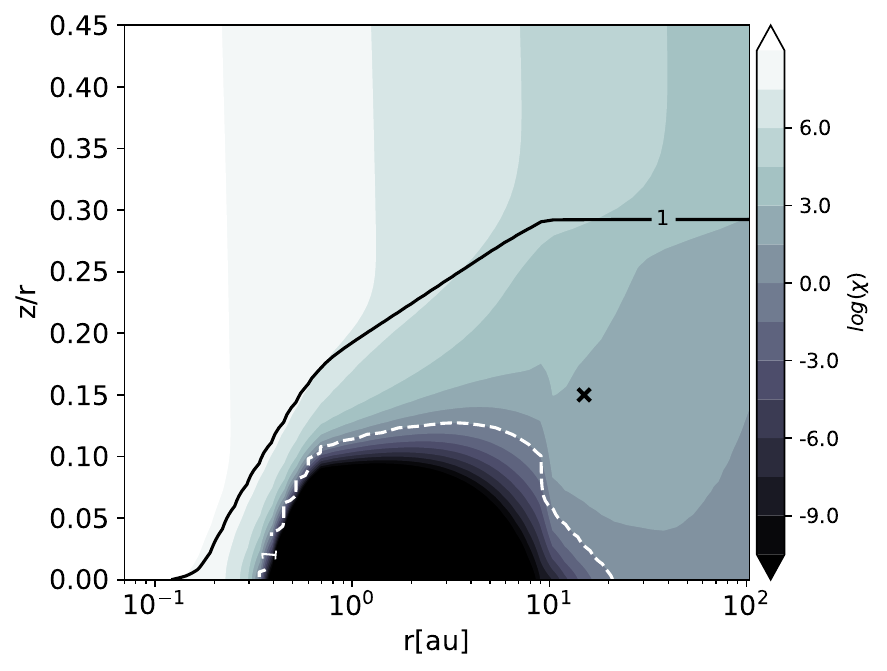}
  \end{minipage}
  \hfill
  \begin{minipage}{0.3\textwidth}
    \centering
    \includegraphics[width=\textwidth]{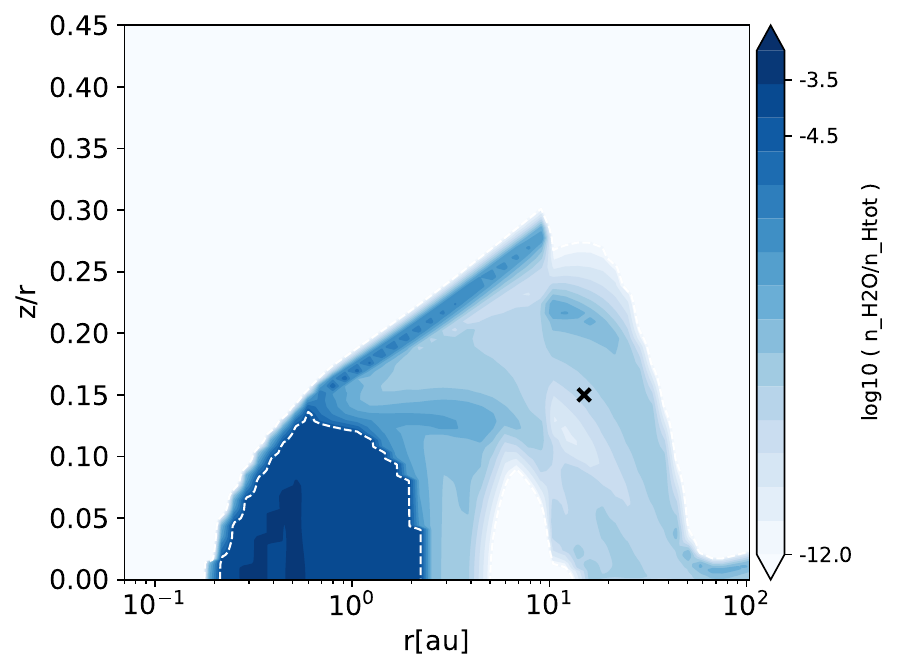}
  \end{minipage}

   \caption{Two-dimensional maps in disk's aspect ratio vs radial distance space for selected disk properties. Top row contains the solution from the best representative model for the XUE 1 disk ($\Go=10^5$); bottom row shows the solution from a non-irradiated disk ($\Go=10^0$) with the same gas and dust composition and structure. Left panels: The grey-scale image shows disk gas temperature with isothermal contours in white dashed lines. The color boxes depict the $15\%-85\%$ emitting areas of the lines that are indicated in the upper-left corner of the figure. Middle panels: strength of the UV radiation field in units of the Draine field---dashed contour indicates the region in the disk that is exposed to an FUV strength equal to the value in the solar neighborhood. Solid contour indicates the region where the radial optical depth is one. Right panels: abundance of gas-phase water relative to the total number of hydrogen nuclei---dashed line encloses $90\%$ of the total abundance. The quantitative analyses presented in Sects. \ref{sec:gas-distribution} and \ref{sec:temperatures-and-abundances} are done at the grid point indicated by the cross.}
  \label{fig:Tgas}
\end{figure*}

The two-dimensional gas temperature from the best representative model is shown in the top-left panel in Fig \ref{fig:Tgas}. The vicinity of the inner radius ($r\lesssim 0.1$ au) is characterized by gas temperatures of $\sim 10^4$ K; similar values are found in the upper disk layers ($z/r\gtrsim 0.5$) at all radii. Close to the midplane, the temperature varies with distance: towards the optically thin dust rim, the temperature drops to $1000$ K; for $0.2 \lesssim r \lesssim 3$ au---where most of the MIRI emission originates---temperature ranges from $1000$ K down to $100$ K; and for $3 \lesssim r \lesssim 10$ au, gas temperature remains below $100$ K. Interestingly, for $r>10$ au, the temperature increases again in response to the external irradiation.

The gas temperature at each grid cell in the computational domain results from balancing out many heating and cooling processes. The final model includes a total of $103$ heating processes and $95$ cooling processes. Notably, FUV photons play an important role in relevant heating mechanisms, such as photoelectric and photochemical heating (e.g., \citealt{Kamp2024}). To quantify the effect of the external irradiation on the temperature, we simulate the XUE 1 disk again, but this time assuming it is not strongly irradiated with $\Go=1.0$. The resulting gas temperature distribution is in the bottom-left panel of Fig. \ref{fig:Tgas}. 

Let us analyze an arbitrary point in the disk, the one with coordinates $\mathcal{P}=(15 \ \mathrm{au}, 0.15)$, which is indicated by the crosses in Fig. \ref{fig:Tgas}. For the irradiated disk, the magnitude of the total heating function at that point is $\Gamma_\mathrm{tot}=7.1 \times 10^{-15} \ \mathrm{erg} \ \mathrm{cm}^{-3} \ \mathrm{s}^{-1}$. For the non-irradiated disk, we find $\Gamma_\mathrm{tot}=1.8 \times 10^{-15} \ \mathrm{erg} \ \mathrm{cm}^{-3} \ \mathrm{s}^{-1}$---this is four times lower than for the irradiated disk. Furthermore, the leading heating mechanism when $\Go=10^5$ is dissociative heating of \ce{H2}; on the contrary, when $\Go=1.0$, the leading heating mechanism is heating by dust thermal accommodation. \rt{Overall, this differences imply a gas temperature at $\mathcal{P}$ of $T_\mathrm{g}=3791$ K (irradiated) and $T_\mathrm{g}=158$ K (non-irradiated). Similarly, the dust temperatures are $T_\mathrm{d}=99$ K (irradiated) and $T_\mathrm{d}=72$ K (non-irradiated).}

Interestingly, points to the right of $\mathcal{P}$ that are at the same height, display higher gas temperatures. Quantitatively, this is explained because the total heating function for those points at larger radii is $\sim 100$ times lower than at $\mathcal{P}$. Also, different heating and cooling mechanisms dominate at each region: dissociative heating of \ce{H2} and Ly-$\alpha$ cooling dominate at $\mathcal{P}$, whereas PAH heating and \ce{O I} line cooling dominate at larger radii. This demonstrates the complex interplay between the heating-cooling balance and the chemistry involved in a self-consistent solution to the equilibrium gas temperature.     

The solution of the continuum radiative transfer determines the strength of the FUV field at each location in the disk. The FUV radiation field (in units of the Draine field) are shown in the middle panels in Fig. \ref{fig:Tgas}. At $\mathcal{P}$, we find $\log{(\chi/\chi_\mathrm{Draine})}=4.7$ for the irradiated disk; for the non-irradiated disk, we find $\log{(\chi/\chi_\mathrm{Draine})}=2.9$ at the same location. We repeat this calculation at a different grid point with coordinates $r=0.3$ au and $z/r=0.1$. This is a characteristic location in the disk from where several of the features observed by MIRI are emitted. At this location, the strength of the FUV field in the irradiated disk is indistinguishable from that in the non-irradiated disk. Clearly, this implies that the FUV field interior to $1$ au is entirely set by the central star.

Given the dependence of the molecular abundances on the gas temperature, it is expected that the external irradiation will affect such abundances in the outer disk. We focus on the gas-phase abundance of water in the XUE 1 disk. Right panels in Fig. \ref{fig:Tgas} display the distribution of water in the irradiated (top) and non-irradiated (bottom) cases. The fractional abundances at $\mathcal{P}$ are $n_{\ce{H2O}}/n_\mathrm{Htot}=7.1 \times 10^{-13}$ and $6.8 \times 10^{-10}$, for the irradiated and non-irradiated cases, respectively ($n_\mathrm{Htot}$ is the total hydrogen number density). At this grid point, the total destruction rate of water (which is equal to the total formation rate because steady state is assumed) is $k^{\mathrm{tot}}_{\ce{H2O}}=4.5 \times 10^{-10} \ \mathrm{cm}^{-3}\ \mathrm{s}^{-1}$ for the irradiated disk, and $k^{\mathrm{tot}}_{\ce{H2O}} = 2.2 \times 10^{-9} \ \mathrm{cm}^{-3}\ \mathrm{s}^{-1}$ for the non-irradiated disk. The model also suggests that the leading destruction mechanism of water in the first case is the neutral-neutral two body reaction: \ce{H + H2O -> OH + H2}. In contrast, for the second case, the leading destruction mechanism is the photo-reaction \ce{H2O + $\gamma_\mathrm{UV}$ -> OH + H}. 

Although the water abundance at $\mathcal{P}$ is lower in the irradiated disk, the right panel in Fig. \ref{fig:Tgas} also suggests an enhanced abundance for lower values of $z/r$, compared to the non-irradiated disk. To quantify this effect, we retrieve from the model the water surface density profile and perform the surface integral to find the mass of water that is contained in the outer disk; i.e., we integrate radially from $10$ au to $100$ au. For the irradiated disk the gas-phase water reservoir is $2.2 \times 10^{-13} \ M_\odot$; for the non-irradiated disk, the gas-phase water reservoir is $4.4\times 10^{-15} \ M_\odot$. Therefore, our model predicts an enhancement by a factor of $100$ in the gas-phase water reservoir in the outskirts relative to a non-irradiated disk. However, the observability of those tracers is expected to be compromised by their low abundance in absolute numbers, due to the depletion effect of the photoevaporating winds. 

\begin{figure*}[h!]  
  
  \centering
  
  \begin{minipage}{0.3\textwidth}
    \centering
    \includegraphics[width=\textwidth]{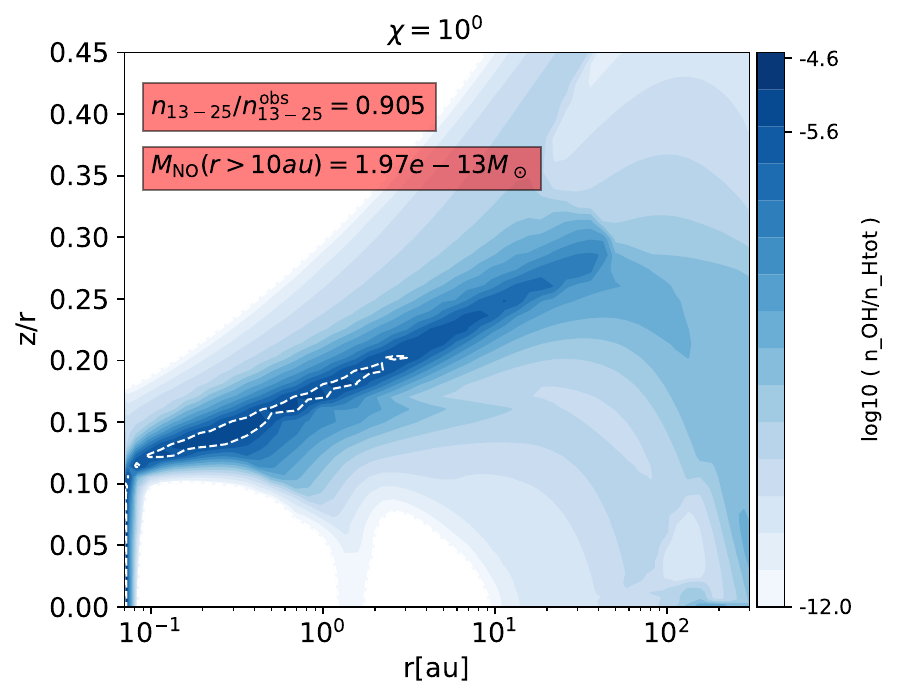}
  \end{minipage}
  \hfill
  \begin{minipage}{0.3\textwidth}
    \centering
    \includegraphics[width=\textwidth]{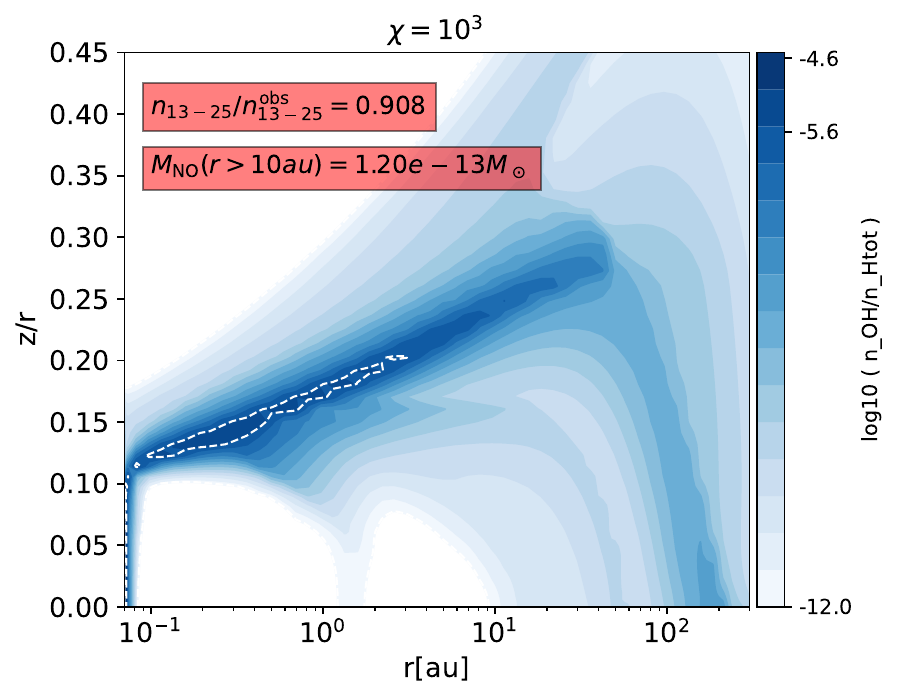}
  \end{minipage}
  \hfill
  \begin{minipage}{0.3\textwidth}
    \centering
    \includegraphics[width=\textwidth]{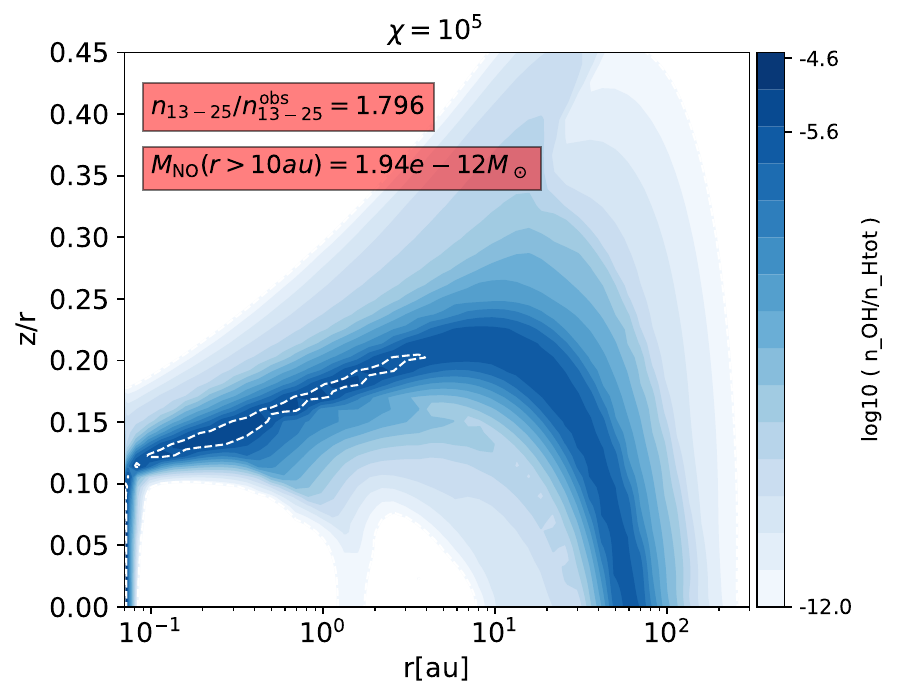}
  \end{minipage}
    
    \vfill
    
  \begin{minipage}{0.3\textwidth}
    \centering
    \includegraphics[width=\textwidth]{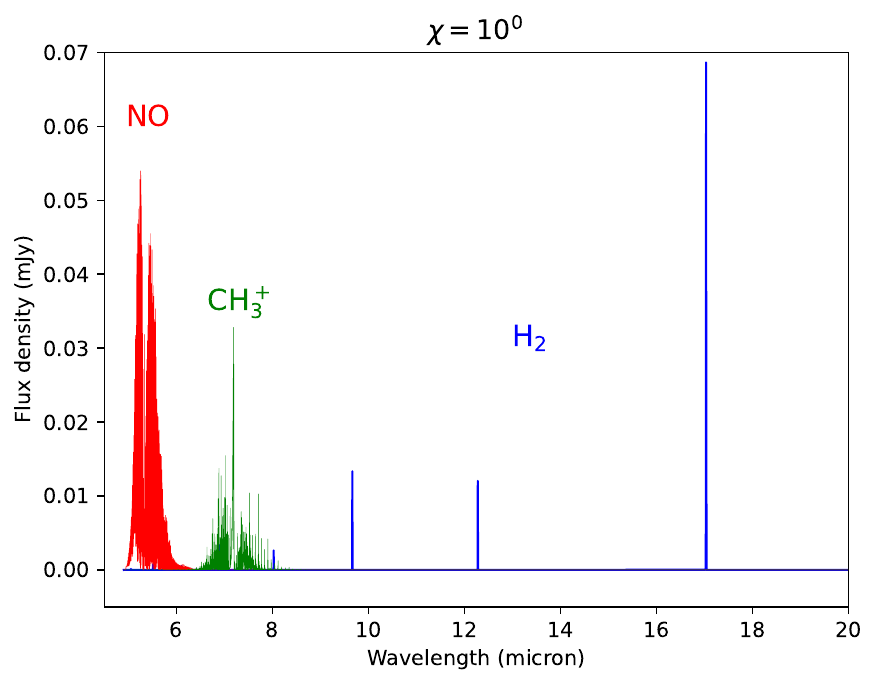}
  \end{minipage}
  \hfill
  \begin{minipage}{0.3\textwidth}
    \centering
    \includegraphics[width=\textwidth]{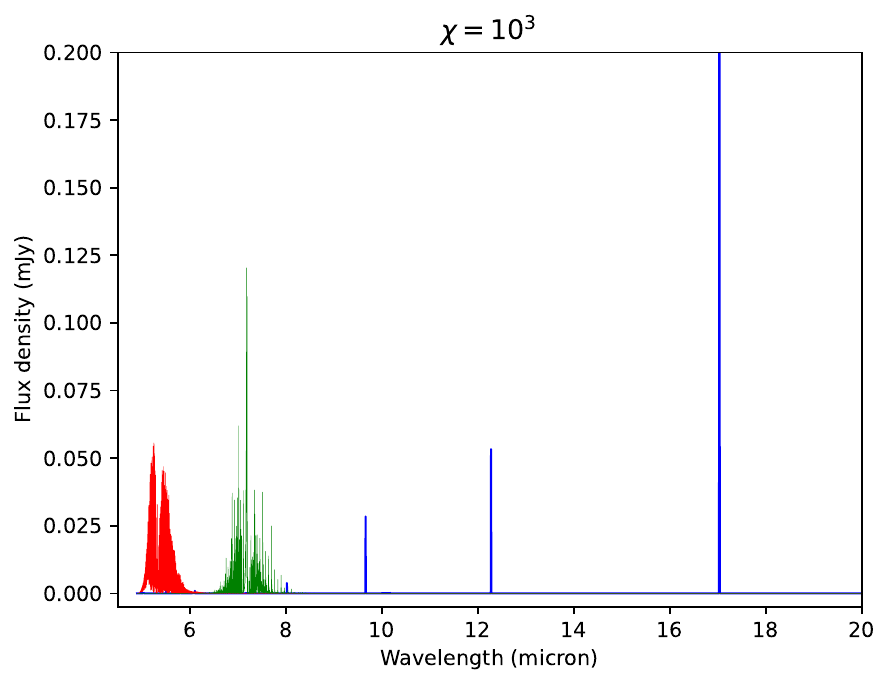}
  \end{minipage}
  \hfill
  \begin{minipage}{0.3\textwidth}
    \centering
    \includegraphics[width=\textwidth]{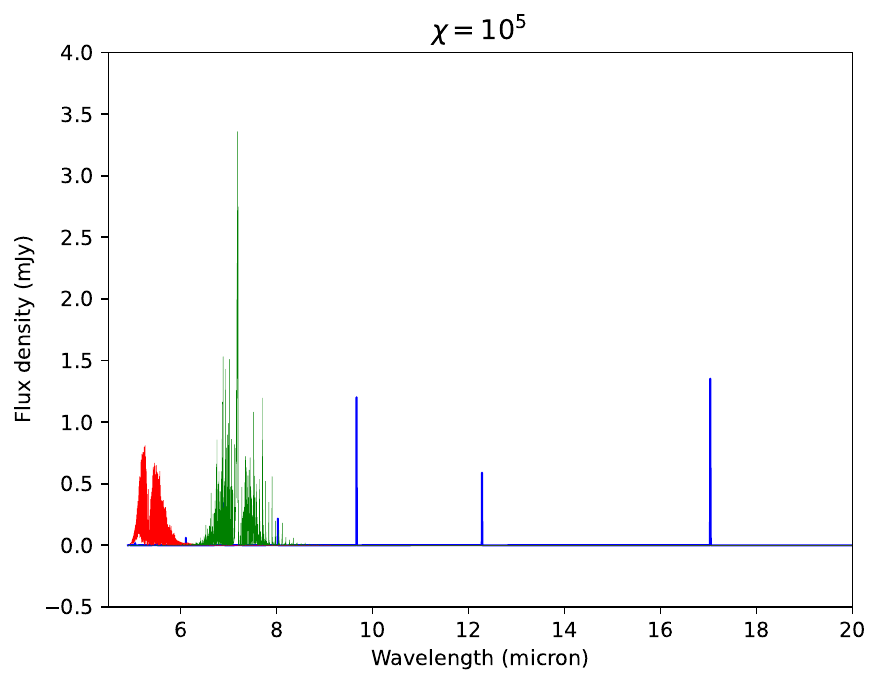}
  \end{minipage}
  \caption{\rt{Effect of external irradiation on the physical structure and line emission of the fiducial disk. \textit{Top:}} Two-dimensional distribution of gas-phase \ce{OH} for different realizations of the fiducial disk. Each realization corresponds to a different value of $\Go$. Dashed lines enclose $90\%$ of the total abundance of \ce{OH}. The insets display for each model the ratio of the synthetic spectral index to the observed one and the corresponding mass of \ce{NO} contained beyond $10$ au. \rt{\textit{Bottom:} Synthetic spectra for \ce{NO, H2} and \ce{CH3+}. Note the difference on the vertical scale.}}
  \label{fig:size-irradiation-degeneracy}
\end{figure*}

\section{Discussion} 
\label{sec:discussion}

\subsection{Emitting regions of mid-infrared lines} 
\label{sec:emitting-regions}
\cite{Ramirez-Tannus2023} reports key physical properties for the XUE 1 disk such as molecular column densities, excitation temperatures, and characteristic sizes of emitting regions, for those lines detected with MIRI. Those properties were derived via slab-modeling the spectra in channels 1 and 3. We perform a similar characterization based on our best representative thermochemical model (Sect. \ref{sec:results}). Note, however, that comparing the results from both approaches must be done with care. While in slab models the retrieval of column densities, temperatures and emitting areas is driven by the data, in thermochemical models those quantities are inferred from the physics that explains the data. Thus, we attempt only a ballpark comparison.
 
We first fetch information about the spatial extent from which a given molecular species emits at a specific wavelength. For a molecule emitting at a certain wavelength, we define its emitting area as the spatial region that encloses $15\%$ to $85\%$ of the line flux, in both the radial and vertical directions (top-left panel in Fig. \ref{fig:Tgas} displays emitting areas of individual lines for a few molecules). Furthermore, the slab models were constrained not by a single line, but by a series of lines spanning a range of wavelengths (see Table 3 in \citealt{Ramirez-Tannus2023}). Consequently, we compute the mean values and standard deviations of each of the aforementioned quantities by averaging the flux-weighted quantity over the spatial and spectral coordinates (i.e., we use Eq. 76 in \citealt{Woitke2024}).     

Results are in Table \ref{table:averaged_quantities} where we also include the values from \cite{Ramirez-Tannus2023}. Physical properties derived from the thermochemical model are in broad agreement with the slab models. Some expected trends---such as larger emitting areas for water lines in channel 3 than in channel 1, as well as a gradient in their excitation temperatures \citep{Banzatti2023}---are seen in both approaches. On the other hand, the largest discrepancy occurs with \ce{CO2}. The slab model retrieves a large emitting area that compensates for a distinctly low column density which falls in the optically thin regime, causing $\log (N)$ and $R$ to become degenerate.

The emitting areas depicted in Fig. \ref{fig:Tgas} suggest that most of the emission observed with MIRI primarily originates in the inner $1$ au of a compact and outer gas-depleted disk. In that inner region the effects of the external irradiation are negligible (Sect. \ref{sec:temperatures-and-abundances}). This explains why the MIRI spectrum of XUE 1 looks so similar to those of nearby, non-irradiated disks.

\begin{deluxetable*}{lccccc|ccc}
\label{table:averaged_quantities}
\tabletypesize{\scriptsize}
\tablewidth{\columnwidth} 
\tablecaption{Physical properties within the emitting regions of mid-IR lines for the XUE 1 disk. The third column lists, for each species, the number of lines used to compute the averaged properties.}
\tablehead{
\colhead{Molecule} & \rt{Vibrational quantum numbers\tablenotemark{\scriptsize a}} & \colhead{No. lines} &\colhead{$\Tave$} &\colhead{log$(\Nave)$} &\colhead{$\Rave$\tablenotemark{\scriptsize b}} &\colhead{$T_\mathrm{ex}$} &\colhead{log$(N)$} &\colhead{$R$}\\
& & & (K) & $(\mathrm{cm}^{-2})$ & (au) & (K) & $(\mathrm{cm}^{-2})$ & (au)\\
& & & \multicolumn{3}{c}{This study} & \multicolumn{3}{c}{\cite{Ramirez-Tannus2023}}} 
\startdata 
\ce{H2O} [7 \ \micron] & \rt{0 1 0 $\rightarrow$ 0 0 0} & 769 & $634 \pm 75$ & $20.0 \pm 0.6$ & 0.33 & 975 & $18.3$ & 0.13 \\
\ce{H2O} [15\ \micron] & \rt{0 0 0 $\rightarrow$ 0 0 0} & 646 & $608 \pm 88$ & $20.0 \pm 0.7$ & 0.38 & 550 & $19.8$ & 0.46 \\
\ce{HCN} & \rt{0 1 1 0 $\rightarrow$ 0 0 0 0} & 211 & $540 \pm 87$ & $19.0 \pm 1.3$ & 0.37 & 575 & $17.3$ & 0.57\\
\ce{C2H2} & \rt{000 0 1 0 1\,\,  u $\rightarrow$ 000 0 0 0+\, g} & 726 & $527 \pm 89$ & $18.3 \pm 2.2$ & 2.7 & 475 & $18.3$ & 0.23\\
\ce{CO2} & \rt{0 1 1 01 $\rightarrow$ 0 0 0 01} & 126 & $574 \pm 147$ & $17.3 \pm 0.4$ & 0.55 & 450 & $14.3$ & 5.30\\
\ce{CO} & \rt{1 $\rightarrow$ 0} & 98 & $1343 \pm 278$ & $19.8 \pm 0.8$ & 0.27  & 2300 & $17.5$ & 0.44\\
\enddata
\tablenotetext{a}{\rt{The quantum numbers are for the vibrational band that contains the strongest transition within a given wavelength range. For \ce{HCN, CO2 and C2H2}, the listed bands include the peaks of the $Q-$branches at $\lambda \sim$ $14.04$, $14.98$ and $13.71\ \micron$, respectively. For \ce{CO}, the band includes the strongest transition of the $P-$branch observable with MIRI in channel 1. For \ce{H2O}, the bands contain the strongest transition in the intervals $5-8 \ \micron$ (rovibrational) and $15-16 \ \micron$ (pure rotational), respectively. Quantum number notation follows HITRAN convention \citep{Gordon2022}}.}\tablenotetext{b}{\rt{Let $\langle R_{15} \rangle$ and $\langle R_{85} \rangle$ be the flux-weighted average distances where the cumulative flux reaches 15\% and 85\%, respectively. We define $\langle R_\mathrm{eff} \rangle$ as the radius of a circle whose area equals that of an annulus with inner and outer radii $\langle R_{15}\rangle$ and $\langle R_{85} \rangle$; this is, $\Rave^2 \equiv \langle R_{85} \rangle^2 - \langle R_{15} \rangle^2$.}}

\end{deluxetable*}

\subsection{Parameter degeneracy: Is the XUE 1 disk truncated by photoevaporation or shielded against FUV photons?}
\label{sec:size-irradiation-degeneracy}
While the parameters listed in Table \ref{tab:model_properties} are derived from physical considerations, those are not the only parameters capable of explaining the observations. This is because our model---just like any other thermochemical model---suffers from multiple degeneracies among its many different parameters (e.g., \citealt{Woitke2015,Cazzoletti2018}). 

There is one degeneracy worth discussing in the context of this work relating the disk's size (parameterized in our model by $\Rtap$) and the strength of the external FUV field (parameterized by $\Go$). This degeneracy implies that the same set of observables could be reproduced if we modeled XUE 1 as a larger disk, but subjected to a weaker external FUV field. In fact, numerical simulations have showed that disks can be shielded from FUV radiation by the residual dust and gas material from the star formation process \citep{Qiao2022,Wilhelm2023}.  

We generate six realizations of the fiducial model (Sect. \ref{sec:modelling}) for different values of $\Go$, from $10^5$ down to $10^0$. Figure \ref{fig:size-irradiation-degeneracy} shows the effect of varying $\Go$ on both $\nspectral/\nspectral^\mathrm{obs}$ and the abundance of \ce{OH} (the precursor of \ce{NO}, see Eq. \ref{eq:NO-chem}), for $\Go=10^0$, $10^3$ and $10^5$. The spectral index reacts to the external irradiation only when $\Go \geq 10^3$. This means that if our second assumption---concerning XUE 1 being constantly exposed to $\Go=10^5$---did not accurately capture the actual irradiation environment of XUE 1, and instead overestimated it by at least a factor of $100$, then the larger fiducial disk could be just as capable of reproducing the observed spectral index as the smaller, strongly irradiated disk. 

Additionally, Fig. \ref{fig:size-irradiation-degeneracy} quantifies how the gas-phase reservoir of \ce{NO} beyond 10 au changes with the strength of the external UV field. For $\Go \leq 10^3$, the mass of \ce{NO} in the outer disk fluctuates closely around a mean value of $1.6 \times 10^{-13} \ M_\odot$. However, for $\Go > 10^3$, the \ce{NO} mass increases substantially, reaching values up to 10 times higher than the mean value at $\Go \leq 10^3$. Again, this means that if our second assumption did not hold, removing gas from the outer disk would not be required to explain the non-detection of \ce{NO}; the solution to the chemistry by itself would lead to a lower abundance of \ce{NO} in the outskirts of the disk.

To break this degeneracy, we need to gather more information on two critical aspects. First, a deeper characterization of the OB star population in the Pismis 24 sub-region is necessary. This includes creating a three-dimensional map for the locations of the ionizing sources relative to XUE 1. This is, however, challenging due to the limitations of current observational facilities. Second, we need better constraints on the local interstellar medium around XUE 1, which involves identifying and characterizing any potential sources of obscuration between the ionizing sources and XUE 1. 

Clearly, this degeneracy could also be resolved in an ensemble sense; this is, by observing and forward modeling a population of irradiated disks, as we did here for XUE 1. Given the large distances to high-mass SFRs, JWST capabilities are key to achieve this goal.     

\rt{In Fig. \ref{fig:size-irradiation-degeneracy}, we also examine the effect of irradiation strength on the synthetic spectra of \ce{NO}, \ce{H2} and \ce{CH3+}. For \ce{NO}, we observe a trend similar to that seen for the continuum slope and \ce{OH} abundance: line fluxes remain comparable for $\chi \leq 10^3$, but increase by more than an order of magnitude when $\chi=10^5$. Both \ce{CH3+} and \ce{H2} exhibit greater sensitivity to external irradiation, with a two-order-of-magnitude increase in flux from $\chi=1$ to $10^5$. Since neither \ce{CH3+} nor \ce{H2} are robustly identified in the XUE 1 spectrum, these findings suggest that a larger disk could potentially explain the observations, but only if $\chi < 10^3$.}


\subsection{Comparison with radiation-hydrodynamical models of photoevaporated disks}
\label{sec:hydro}
Our best representative model suggests a compact and low-mass disk around XUE 1. Truncation of disks is expected in high FUV environments \citep[e.g.,][]{Clarke2007, 2022MNRAS.514.2315C}; however, our calculations impose a depleted hydrostatic profile beyond $10$ au which might not necessarily reflect the structure of a photoevaporative wind as a hydrodynamic flow. We check this by comparing with bespoke radiation hydrodynamic simulations of the wind for disk parameters suggested by our best representative model, calculated using the \torus \ code \citep{Bisbas2015,Bisbas2012,Harries2019}. \torus \ performs photodissociation region and hydrodynamic calculations iteratively to solve for the steady state wind structure. These models are extensively used in the \textsc{fried} (FUV Radiation Induced Evaporation of Discs) grids \citep[see][for further details]{Haworth2018, 2023MNRAS.526.4315H}. These calculations are FUV-only and so only apply to the wind interior to the ionisation front, however that is sufficient here given our main focus is on the surface density profile close to the disk outer edge. We retrospectively calculated how deep into the FUV-only model ionising radiation would penetrate by solving ionisation equilibrium on a cell by-cell basis assuming hydrogen only gas. Geometrically diluting the ionising radiation from the UV sources in the cluster yields an ionising flux at XUE 1 of $1.59\times10^{13}$\,photons\,cm$^{-2}$\,s$^{-1}$, in which case the ionisation front only makes it 0.1\,au into the grid. That estimate also does not account for material beyond the grid edge, nor the fact that the incident ionising flux may be further attenuated by the true separation from the UV sources being larger than the projected, as well as dust absorption.



Our model for the XUE 1 disk (Sect. \ref{sec:results}) serves as boundary condition for the \torus \ simulation. Accordingly, we adopt a value of $10$ au as the radius of the \torus \ disk; that is, the $\Rtap$ parameter of the \prodimo \ model is taken as the inner boundary for the photoevaporative flow in the \torus \ model. 

The rationale behind our choice of $\Rtap$ as the flow's inner boundary is twofold. First, our choice can be heuristically motivated by the necessity to remove gas beyond $10$ au to explain the non-detection of \ce{NO} (Sect. \ref{sec:gas-distribution}). Second, \prodimo \ predicts values above $1000$ K for the gas temperature for $r>10$ au (see Fig. \ref{fig:Tgas})---these temperature values are comparable to the threshold $T_\mathrm{g} \lesssim 3000$ K that \torus \ assumes when labeling a given cell as a PDR \citep{Bisbas2015}.

 The gas density structure for $r<10$ au is kept fixed to the value that explains the JWST/MIRI data. The initial surface density profile is modeled as $\Sigma(r,t=0)=\Sigma_\mathrm{1 au} (r/\mathrm{au})^{-1}$, with $\Sigma_\mathrm{1 au}=10 \ \mathrm{g} \ \mathrm{cm}^{-2}$, which is also informed by the \prodimo \ model (see Fig. \ref{fig:dust-gas-distribution}). For the flow, we adopt a gas-to-dust ratio of $10^4$---this means that we treat the dust population in the disk as an evolved one. This treatment aligns with our adoption of a dust size distribution for the disk, where grains as big as $3$ mm are allowed to be present. In view of such size distribution ruling the disk opacity, and considering that grains larger than $0.1 \ \micron$ are not expected to be entrained in the flow \citep{Facchini2016}, it is logical to assume that most of the dust mass will remain in the disk, which implies a dust-depleted flow. We use a mean grain cross section of $\sigma_\mathrm{grain}=5.5 \times 10^{-23} \ \mathrm{cm}^{-2}$ to set the opacity in the flow.   

We let the simulation run until the flow structure achieves steady state. The resulting gas surface density is depicted in Fig. \ref{fig:dust-gas-distribution} (black curve). The steady-state solution for $r>10$ au is in good agreement with the parameterised density structure from \prodimo. This suggests that, for this particular combination of star, disk, and FUV field parameters, the inner wind behaves more like a hydrostatic atmosphere attached to the disk, which explains why this approximation is effective. 

The \torus \  calculation yields a mass loss rate $\dot{M}_\mathrm{wind}=5 \times 10^{-10} \ M_\odot \ \mathrm{yr}^{-1}$. This value combined with the inferred disk mass (Sect. \ref{sec:gas-distribution}) implies a depletion timescale $\tau_\mathrm{depletion} \equiv M_\mathrm{disk}/\dot{M}_\mathrm{wind} \approx 0.12$ Myr, which is similar to the values inferred from proplyds in the Orion Nebular Cluster \citep[e.g.][]{1999AJ....118.2350H}.       

Accounting for both non-LTE effects in the thermochemical solution of \ce{NO} (Sect. \ref{sec:gas-distribution}) and potential shielding mechanisms against FUV photons (Sect. \ref{sec:size-irradiation-degeneracy}) may increase our disk mass estimate. Consequently, the wind depletion timescale derived in this section should be interpreted as a lower limit.

\section{Conclusions} 
\label{sec:conclusions}
In this work, we analyze JWST/MIRI spectroscopic and archival photometric data of the externally irradiated protoplanetary disk around the T Tauri star XUE 1. The XUE 1 disk is part of NGC 6357, a high-mass star forming region (SFR) with a UV radiation environment that surpasses those in nearby SFRs due to its higher density of O type stars. Our analysis is conducted by building a thermochemical model for the disk that is capable of self-consistently explaining the available set of observations: $V$ and $Z$ bands HST/ACS photometry; $J,H,K_s$ VISTA/NIR and [$3.6 \ \micron$], [$4.5 \ \micron$], [$5.8 \ \micron$], and [$8.0 \ \micron$] Spitzer/IRAC upper flux limits; JWST/MIRI continuum; and the JWST/MIRI molecular emission across channels 1 and 3. We infer from this model one set of suitable parameters that characterize the disk's physical structure. Our main conclusions are:

\begin{enumerate}
    
    \item We confirm the presence of amorphous ($\silica$, $\pyroxene$, and $\olivine$) and crystalline (forsterite and enstatite) silicate grains populating the disk surface. We find no evidence supporting the presence of troilite or fayalite (Sect. \ref{sec:continuum-characterisation}).
    
    
    \item Given our model assumptions, we rule out a disk with a dust reservoir of $M_\mathrm{dust} \geq 10^{-6} \ M_\odot$, that is radially extended ($\Rtap \geq 100$ au) and externally irradiated with $\Go = 10^5$ (the FUV-integrated energy density relative to the solar neighborhood value). We find that such physical structure would be inconsistent with the available data (Sect. \ref{sec:dust-distribution}).

    \item With the caveat that none of the currently available data are a probe for the disk mass, our model suggests a dust mass of $0.3 \ \Mearth$ and a gas mass of $6 \times 10^{-5} \ M_\odot$. The model also suggests an outer disk ($r>10$ au) that is gas-depleted. Our inferred dust and gas distributions are such that the gas-to-dust ratio is: constant and equal to $1$ for $r>10$ au; constant and equal to $100$ for $1<r<10$ au; and variable, with an integrated value of $167$, for $0.07<r<1$ au (Sects. \ref{sec:dust-distribution} and \ref{sec:gas-distribution}). 
    
    \item \rt{The external irradiation sets the disk FUV field beyond $10$ au and stellar photons dominate the FUV field at shorter radii. Beyond $10$ au, gas temperature reaches values above $1000$ K---an order of magnitude higher than the characteristic temperatures expected in a non-irradiated disk with similar physical structure. Similarly, the effect of the external irradiation on molecular abundances is only noticeable beyond $10$ au. In particular, for gas-phase water, we predict a 100-fold increase in abundance compared to a non-irradiated disk with the same physical structure (Sect. \ref{sec:temperatures-and-abundances}).}  

\end{enumerate}

The unrivaled capabilities of JWST, combined with state-of-the-art numerical codes, enable us to present the first data-informed thermochemical model of an extremely irradiated disk at kiloparsec-distance scales. Our results imply that dust and gas material can be present within the first $10$ au in a strongly irradiated disk. This supports the idea that planet formation is possible even in environments with extreme UV irradiation. 

While the physical structure of XUE 1 aligns qualitatively with an external photoevaporation scenario, we cannot rule out model degeneracies that could allow for a larger, more massive disk under conditions of reduced external irradiation. Observing and characterizing a population of highly irradiated disks will help to break this degeneracy. A synergistic approach with other facilities, particularly at sub-millimeter wavelengths, is also needed to improve the characterization of the XUE 1 and other highly irradiated disks. This work serves as an initial foundation for such future efforts.


\newpage
We are grateful to the anonymous referee for providing thoughtful and helpful comments that improved the manuscript. We thank Jayatee Kanwar and Sebastián Hernández for insightful discussions on modeling mid-IR spectroscopy, and Rebecca Zoshak for proofreading the manuscript. This project is supported by NASA STScI grant JWST-GO-01759.002-A. The Center for Exoplanets and Habitable Worlds is supported by the Pennsylvania State University and the Eberly College of Science. The research of T.P.~was partly supported by the Deutsche Forschungsgemeinschaft (DFG, German Research Foundation) in the research unit \textit{Transition Discs} (Ref no.~325594231 -FOR~2634/2 - TE 1024/2-1) and in the  Excellence Cluster \textit{ORIGINS} - EXC 2094 - 390783311. E.S. is supported by the international Gemini Observatory, a program of NSF NOIRLab, which is managed by the Association of Universities for Research in Astronomy (AURA) under a cooperative agreement with the U.S. National Science Foundation, on behalf of the Gemini partnership of Argentina, Brazil, Canada, Chile, the Republic of Korea, and the United States of America. TJH acknowledges funding from a Royal Society Dorothy Hodgkin Fellowship and UKRI guaranteed funding for a Horizon Europe ERC consolidator grant (EP/Y024710/1). A.B. and J.F. acknowledge support from the Swedish National Space Agency (grant No. 2022-00154). MCRT acknowledges support by the German Aerospace Center (DLR) and the Federal Ministry for Economic Affairs and Energy (BMWi) through program 50OR2314 ‘Physics and Chemistry of Planet-forming disks in extreme environments’. V.R. acknowledges the support of the European Union’s Horizon 2020 research and innovation program and the European Research Council via the ERC Synergy Grant “ECOGAL” (project ID 855130). This research has made use of the SVO Filter Profile Service ``Carlos Rodrigo", funded by MCIN/AEI/10.13039/501100011033/ through grant PID2020-112949GB-I00. \rt{Some/all of the data presented in this article were obtained from the Mikulski Archive for Space Telescopes (MAST) at the Space Telescope Science Institute. The specific observations analyzed can be accessed via \dataset[doi:10.17909/6nwv-6f56]{https://archive.stsci.edu/doi/resolve/resolve.html?doi=10.17909/6nwv-6f56}}

%

\vspace{5mm}
\facilities{JWST (MIRI-MRS)}






\bibliography{sample631}{}
\bibliographystyle{aasjournal}



\end{document}